\documentclass[sigplan,screen]{acmart}
\copyrightyear{2026}
\acmYear{2026}
\setcopyright{cc}
\setcctype{by}
\acmConference[ASPLOS '26]{Proceedings of the 31st ACM International Conference on Architectural Support for Programming Languages and Operating Systems, Volume 2}{March 22--26, 2026}{Pittsburgh, PA, USA}
\acmBooktitle{Proceedings of the 31st ACM International Conference on Architectural Support for Programming Languages and Operating Systems, Volume 2 (ASPLOS '26), March 22--26, 2026, Pittsburgh, PA, USA}
\acmPrice{}
\acmDOI{10.1145/3779212.3790147}
\acmISBN{979-8-4007-2359-9/2026/03}
\makeatletter
\let\@ACM@copyright@check@cc\relax
\makeatother

\AtBeginDocument{%
  }

\usepackage{multirow}
\usepackage{bbding}
\usepackage{bm}
\usepackage{pifont}
\usepackage{amsmath}
\usepackage{xcolor}
\usepackage{colortbl}
\usepackage{booktabs}
\usepackage{tabularx} 
\usepackage{makecell} 
\usepackage{footnote}
\usepackage{tablefootnote}
\usepackage{float}
\usepackage{threeparttable}
\usepackage{textcomp}

\newcommand{\xietong}[1]{{\color{black}#1}}

\newcommand{\method}{CREATE}

\begin{document}


\title[CREATE: Cross-Layer Resilience Characterization and Optimization for Embodied AI Systems]{\method: \underline Cross-Layer Resilience Characterization and Optimization for Efficient yet \underline Reliable \underline Embodied \underline AI Sys\underline{te}ms}



\author{Tong Xie}
\orcid{0009-0007-1259-3762}
\affiliation{%
  \institution{School of Integrated Circuits}
 \institution{Peking University}
  \city{Beijing}
  \country{China}
}
\email{xietong@stu.pku.edu.cn}

\author{Yijiahao Qi}
\orcid{0009-0006-7649-1127}
\affiliation{%
  \institution{School of EECS}
 \institution{Peking University}
  \city{Beijing}
  \country{China}
}
\email{qiyijiahao@stu.pku.edu.cn}

\author{Jinqi Wen}
\orcid{0009-0002-0025-6439}
\affiliation{%
  \institution{School of EECS}
 \institution{Peking University}
  \city{Beijing}
  \country{China}
}
\email{2200012451@stu.pku.edu.cn}

\author{Zishen Wan}
\orcid{0000-0002-2982-5351}
\affiliation{%
 \institution{Georgia Institute of Technology}
  \city{Atlanta}
  \state{GA}
  \country{USA}
}
\email{zishenwan@gatech.edu}

\author{Yanchi Dong}
\orcid{0009-0004-5321-8297}
\affiliation{%
  \institution{School of Integrated Circuits}
 \institution{Peking University}
  \city{Beijing}
  \country{China}
}
\email{dongyanchi@pku.edu.cn}

\author{Zihao Wang}
\orcid{0000-0001-8396-3707}
\affiliation{%
  \institution{Institute for Artificial Intelligence}
 \institution{Peking University}
  \city{Beijing}
  \country{China}
}
\email{zhwang@stu.pku.edu.cn}

\author{Shaofei Cai}
\orcid{0000-0002-1195-7276}
\affiliation{%
  \institution{Institute for Artificial Intelligence}
 \institution{Peking University}
  \city{Beijing}
  \country{China}
}
\email{caishaofei@stu.pku.edu.cn}

\author{Yitao Liang}
\orcid{0009-0009-0944-7814}
\affiliation{%
  \institution{Institute for Artificial Intelligence}
 \institution{Peking University}
  \city{Beijing}
  \country{China}
}
\email{yitaol@pku.edu.cn}

\author{Tianyu Jia}
\orcid{0000-0002-4570-4613}
\affiliation{%
  \institution{School of Integrated Circuits}
 \institution{Peking University}
  \city{Beijing}
  \country{China}
}
\email{tianyuj@pku.edu.cn}

\author{Yuan Wang}
\orcid{0000-0002-4951-4286}
\affiliation{%
  \institution{School of Integrated Circuits}
 \institution{Peking University}
  \city{Beijing}
  \country{China}
}
\email{wangyuan@pku.edu.cn}

\author{Runsheng Wang}
\orcid{0000-0002-7514-0767}
\affiliation{%
  \institution{School of Integrated Circuits}
 \institution{Peking University}
  \city{Beijing}
  \country{China}
}
\email{r.wang@pku.edu.cn}

\author{Meng Li}
\orcid{0000-0002-7212-2264}
\affiliation{%
  \institution{Institute for Artificial Intelligence}
 \institution{Peking University}
  \city{Beijing}
  \country{China}
}
\email{meng.li@pku.edu.cn}
\authornote{Corresponding author}

\renewcommand{\shortauthors}{Tong Xie et al.}




\begin{abstract}

Embodied Artificial Intelligence (AI) has recently attracted significant attention as it bridges AI with the physical world. Modern embodied AI systems often combine a Large Language Model (LLM)-based planner for high-level task planning and a reinforcement learning (RL)-based controller for low-level action generation, enabling embodied agents to tackle complex tasks in real-world environments. 
However, deploying embodied agents remains challenging due to their high computation requirements, especially for battery-powered local devices. Although techniques like lowering operating voltage can improve energy efficiency, they can introduce bit errors and result in task failures.

\xietong{
In this work, we propose~\method, a general design principle that leverages heterogeneous resilience at different layers for synergistic energy-reliability co-optimization.}
For the first time, we conduct a comprehensive error injection study on modern embodied AI systems and observe an inherent but heterogeneous fault tolerance.
Building upon these insights, we develop an \textit{anomaly detection and clearance} mechanism at the circuit level to eliminate {outlier errors}. At the model level, we propose a \textit{weight-rotation-enhanced planning} algorithm to improve the fault tolerance of the LLM-based planner. 
Furthermore, we introduce an application-level technique, \textit{autonomy-adaptive voltage scaling}, to dynamically adjust the operating voltage of the controllers. 
The voltage scaling circuit is co-designed to enable online voltage adjustment.
Extensive experiments demonstrate that without compromising task quality, \method~achieves 40.6\% computational energy savings on average over nominal-voltage baselines and 35.0\% over prior-art techniques.
\xietong{This further leads to 29.5\% to 37.3\% chip-level energy savings and approximately a 15\% to 30\% improvement in battery life.
}
\vspace{-5pt}
\end{abstract}

\maketitle
\vspace{-9pt}
\section{Introduction}
\label{sec: intro}

\begin{figure}[!tb]
    \centering
    \includegraphics[width=1\linewidth]{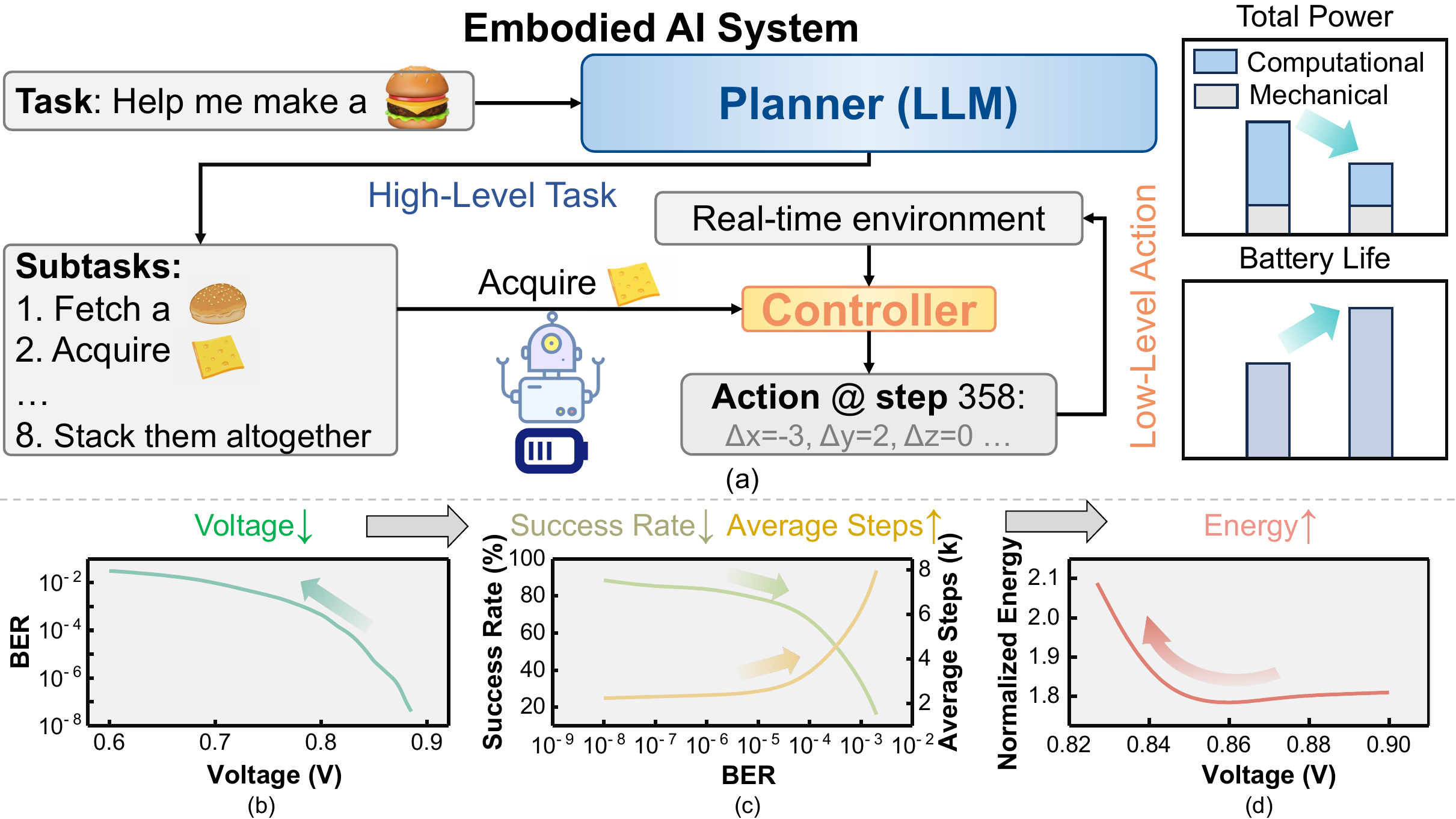}
    \vspace{-15pt}
    \caption{
    \textbf{Embodied AI System Overview.}
    (a) Embodied AI systems employ a hierarchical paradigm to tackle complex tasks: an LLM-based planner decomposes a task (often described in texts) into subtasks; then for each subtask, a low-level controller generates detailed actions step by step. These systems are typically battery-powered. (b) Lowering operating voltage introduces bit errors, causing (c) significant task performance degradation (e.g., lower task success rates and more task execution steps). (d) This in turn leads to higher energy consumption per task. 
    }
    \label{fig: figure 1}
    \vspace{-14pt}
\end{figure}

Embodied artificial intelligence (AI) is a class of intelligent systems capable of understanding, reasoning, and interacting with the physical world \cite{liu2024aligning,duan2022survey,xu2024survey}.
Distinguished by their proficiency in executing complex long-horizon, multi-objective tasks, these systems have demonstrated broad application potential across various domains, including industrial manufacturing \cite{brohan2023rt}, logistics and transportation \cite{gan2022threedworld}, and search-and-rescue operations \cite{zhou2024hazard}.
As {shown} in Fig. \ref{fig: figure 1}(a), to achieve such capabilities, embodied agents typically employ large language models (LLMs) or multi-modal language models (MLMs) as \textit{planners} to decompose the entire \textit{tasks} into specific \textit{subtasks}, alongside {reinforcement learning (RL)-based} \textit{controllers} to determine low-level actions for each subtask, which typically requires thousands of \textit{steps} \cite{ahn2022can, driess2023palm, song2023llm, wang2023voyager, wang2024jarvis, black2024pi_0, brohan2023rt}. Task performance is usually measured by the success rate and average completion steps \cite{wang2024jarvis,huang2025dadu,song2023llm, brohan2023rt,wang2023voyager, black2024pi_0}. 




Unlike traditional robotic systems driven by predefined rules \cite{suleiman2019navion, gan2021eudoxus}, embodied agents leverage large-scale LLMs, which would often consume trillions of operations per inference. Although cloud-based inference via APIs enables access to giant LLMs, e.g., GPT-4 \cite{achiam2023gpt}, it 
introduces high communication latency and reliance on Wi-Fi, limiting desirable real-time interaction capabilities for many applications. Therefore, recent research has begun to explore the end-to-end deployment of embodied agents on edge hardware, such as GPUs \cite{wen2025tinyvla, brohan2022rt, driess2023palm} and customized accelerators \cite{wan2021survey, ahn2022can, huang2025dadu, brohan2023rt, cuhk2023airship}, such as TPU-like systolic arrays \cite{jouppi2017datacenter, markidis2018nvidia}. 

\begin{figure*}[!tb]
    \centering
    \includegraphics[width=1\linewidth]{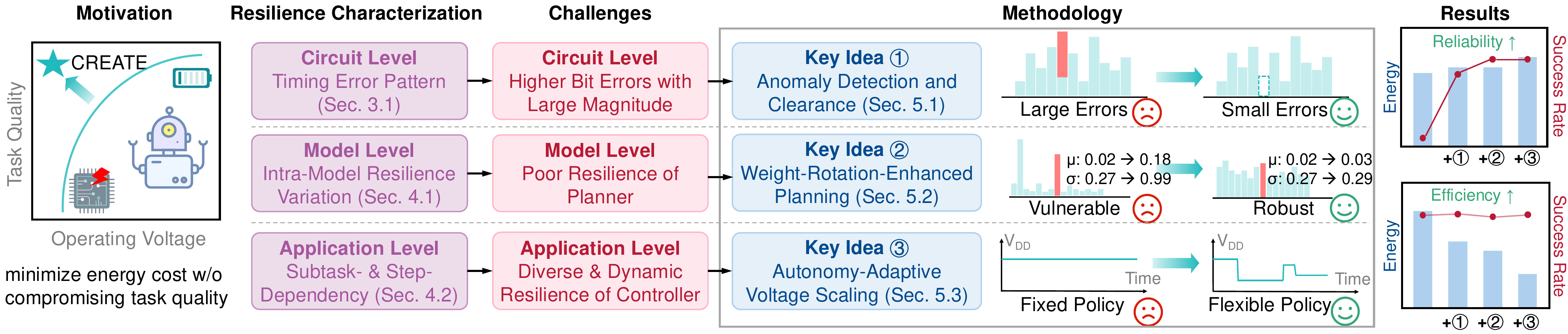}
    \vspace{-16pt}
    \caption{\textbf{\method~Overview.} \method~is a
    \xietong{general design principle}
    to enable efficient yet reliable embodied AI systems. At the \underline{circuit} level, \textit{anomaly detection and clearance} mitigate large timing errors. At the \underline{model} level, \textit{weight-rotation-enhanced planning} redistributes LLM activations to improve robustness. At the \underline{application} level, \textit{autonomy-adaptive voltage scaling} dynamically adjusts voltage based on task demands. These techniques collectively boost both reliability and efficiency. }
    \label{fig: Overview}
\end{figure*}


However, such edge deployment of embodied AI systems is often strictly energy-constrained due to reliance on battery power 
\cite{liu2024aligning, mcnulty2022review, wan2021survey, cuhk2023airship,figureai2025helix}.
\xietong{With intensive computational workloads (e.g., long-context LLM inference and high-frame-rate image processing), these computation platforms have been extensively reported to consume substantial power
\cite{wan2025reca, kim2024openvla,brohan2023rt, driess2023palm, li2023vision, huang2025dadu, ting2025hiper, suleiman2019navion, yue2024deer, jabbour2024generative}.
For example, \cite{jabbour2024generative} reported that the {computing} platform accounted for over 50\% of total power on quadruped robots \cite{anybotics2023anymal}, and \cite{huang2025dadu} demonstrated that LLM inference dominated the total energy {consumption} in a representative robotic setup using an NVIDIA GPU and Franka robot arm \cite{gaz2019dynamic, li2023vision, kim2024openvla}. Therefore, it is imperative to improve the power efficiency of embodied AI systems. Given the quadratic relation between {power consumption} and operating voltage, directly reducing the voltage can improve processing efficiency.} {This method}, however, can {induce} timing violations \cite{dixit2021silent, moghaddasi2023dependable, jiao2017clim, huang2017variability} and increase computation bit error rates (BERs), as shown in Fig. \ref{fig: figure 1}(b), leading to severe degradation of task performance, such as lower task success rates and more execution steps (Fig. \ref{fig: figure 1}(c)). These can ultimately increase the energy consumption to complete a single task (Fig. \ref{fig: figure 1}(d)).


Various techniques have been proposed to balance reliability with efficiency at the circuit, algorithm, and system levels \xietong{\cite{libano2018selective, khoshavi2020shieldenn, talpes2020compute, zhang2018thundervolt, zhang2019fault,pandey2019greentpu, ernst2003razor, gundi2020effort, whatmough2018dnn,huang1984algorithm, bal2023novel, safarpour2021algorithm, xue2023approxabft,schorn2019efficient, kim2018matic, he2019noise,wan2024mulberry,jha2022exploiting, wan2024vulnerability, shah2024characterizing}}, but all remain prohibitively costly or impractical for embodied AI systems. 
Unlike traditional robotics, modern embodied agents are predominantly based on AI and deep neural networks (DNNs), which exhibit inherent error resilience \cite{li2017understanding, reagen2018ares, mahmoud2021optimizing, wan2021analyzing, agarwal2023resilience, wan2024mulberry, xie2025realm}.
For example, as in Fig.~\ref{fig: figure 1}(c), task success rates (\texttt{stone} in \cite{wang2024jarvis}) do not immediately degrade under low BERs, suggesting opportunities to save energy by strategically reducing the operating voltage without compromising the task performance. However, 
the error resilience of embodied AI systems is not well characterized yet, making it hard to determine a proper voltage scaling policy. While prior studies have examined the resilience of individual DNNs (e.g., convolutional neural networks (CNNs), RL algorithms, etc) \cite{li2017understanding,reagen2018ares,mahmoud2021optimizing,wan2021analyzing} and traditional robotic systems \cite{wan2024vulnerability, shah2024characterizing}, the complex collaborations among diverse models as well as the sophisticated applications in embodied AI systems can produce fundamentally different resilience behaviors.



Therefore, to enable efficient yet reliable embodied AI systems through voltage scaling, for the first time, we perform large-scale error injection experiments to systematically characterize their error resilience. Our investigation reveals inherent but heterogeneous system robustness: while both the planner and the controller show good error robustness at low BERs (e.g., $\leq10^{-7}$), the controller exhibits much higher resilience at higher BERs. (e.g., $10^{-7}$ to $10^{-3}$). 
We also analyze the resilience variation of
different network components in the planner and controller, as well as 
dependency on subtasks and execution status. 
We observe that the systematic activation outliers in the LLM, when combined with the subsequent normalization operations, are the root cause of the planner's poor resilience at higher BERs. Meanwhile, the controller exhibits varying resilience patterns across different subtasks and action steps during task execution.

Building on these insights, we propose \method, a \xietong{general design principle that synergistically integrates optimizations across circuit, model, and application levels to work progressively}, as illustrated in Fig. \ref{fig: Overview}. At the \textbf{circuit level}, we introduce an \textit{anomaly detection and clearance} mechanism to suppress large errors induced by timing violations, \xietong{which serves as a solid foundation for subsequent optimizations although conceptually simple}. However, smaller yet still critical errors, arising from the activation distributions of the LLM planner, persist and degrade performance.
To address this, at the \textbf{model level}, we propose \textit{weight-rotation-enhanced planning} to redistribute activation patterns of the LLM planner and improve robustness. To further optimize energy efficiency \xietong{of the controller}, at the \textbf{application level},
we propose \textit{autonomy-adaptive voltage scaling}, which employs a flexible voltage control policy based on the current subtask execution status. We further customize a circuit for dynamic voltage scaling in systolic arrays and low-dropout regulators (LDOs) to holistically implement these optimizations. 
Our contributions can be summarized as follows:

\vspace{-8pt}
\begin{itemize}
\item We conduct a comprehensive error injection study on representative embodied AI systems to analyze the impact of voltage-scaling-induced errors on task performance. Our analysis reveals that the controller demonstrates significantly better error resilience than the planner. We further investigate the resilience behaviors across different network components and execution steps.
To the best of our knowledge, this is the first comprehensive study of embodied AI system's error resilience. 
\item We introduce \method, \xietong{a general design principle that exploits the heterogeneous error resilience of embodied AI systems with cross-layer optimizations} to minimize energy consumption without compromising their reliability and task performance. This framework is implemented via a customized circuit and LDOs.
\item Extensive experiments demonstrate \method~ significantly improves error tolerance and energy efficiency while achieving generality across diverse tasks and voltage conditions. 
By aggressively lowering the operating voltage, \method~achieves an average 40.6\% improvement in \xietong{computational} energy efficiency over the nominal-voltage operation and 35.0\% over state-of-the-art (SOTA) methods, \xietong{translating to 29.5\% to 37.3\% chip-level energy savings and about 15\% to 30\% longer battery life, all with iso-task quality.} 
\end{itemize}

\vspace{-9pt}
\section{Background}
\label{sec: background}

\begin{table*}[t]
\centering
\footnotesize
\caption{Representative Embodied Agent Systems that \method~Can Support}
\vspace{-7pt}
\label{tab:embodied}
\renewcommand*{\arraystretch}{1.05}
\resizebox{\linewidth}{!}{
\begin{tabular}{c|cc|cc|ccc}
\hline \hline
\textbf{Embodied} & \multicolumn{2}{c|}{{\textbf{Planner}}} & \multicolumn{2}{c|}{{\textbf{Controller}}} & \multicolumn{3}{c}{{\textbf{Tasks}}} \\
\cline{2-8}
\textbf{System} & {Model} & {\#Params (B)} & {Model} & {Architecture} & {Description} & {Example} & {Benchmark} \\
\hline \hline

\textbf{SayCan} \cite{ahn2022can} & PaLM \cite{chowdhery2023palm} & 540 & MT-Opt \cite{kalashnikov2021mt} & CNN & navigation  & "move to the table" & ALFRED \cite{shridhar2020alfred} \\
\hline

\textbf{PaLM-E} \cite{driess2023palm}& PaLM-E \cite{driess2023palm} & 562 & MT-Opt  \cite{kalashnikov2021mt} & CNN & manipulation  & "place water bottle upright"& Language-Table \cite{lynch2023interactive} \\ \hline

\textbf{LLM-Planner} \cite{song2023llm} & GPT-3 \cite{brown2020language}& 175 & HLSM  \cite{blukis2022persistent} & Transformer & navigation  & "move to the table" & ALFRED \cite{shridhar2020alfred} \\
\hline

\textbf{RT-2} \cite{brohan2023rt} & PaLI-X  \cite{chen2023pali} & 55 & RT-1 \cite{brohan2022rt} & Transformer &  manipulation & "place water bottle upright" &Language-Table \cite{lynch2023interactive} \\ \hline

\textbf{Voyager} \cite{wang2023voyager}& GPT-4 \cite{achiam2023gpt} & 1,760 & - & - & multi-task & "obtain an iron sword" & Minecraft \cite{guss2019minerl}\\ \hline

\textbf{OpenVLA} \cite{kim2024openvla} & LLaMA  \cite{liu2023visual} & 7 &- &- & mult-task &"put bottle on top of cabinet"& LIBERO \cite{liu2023libero} \\ \hline

\textbf{RoboFlamingo} \cite{li2023vision} & OpenFlamingo  \cite{awadalla2023openflamingo} & 3 & -&- & mult-task &"pull handle to open drawer"& CALVIN \cite{mees2022calvin}\\ \hline

\textbf{Octo} \cite{ghosh2024octo} & - & - & Octo \cite{ghosh2024octo}&Transformer& mult-task &"bring me 3 different sodas"& OXE \cite{o2024open} \\ \hline

\textbf{LEO} \cite{huang2024embodied} & Vicuna \cite{chiang2023vicuna} & 7 & ScanRefer \cite{chen2020scanrefer}&Transformer&navigation &"find the toilet"&HM3D \cite{ramakrishnan2021hm3d} \\ \hline


\rowcolor{gray!25}
\textbf{JARVIS-1} \cite{wang2024jarvis} & LLaVA  \cite{liu2023visual} & 8 & STEVE-1 \cite{lifshitz2023steve}&Transformer&mult-task &"obtain an iron sword"&Minecraft \cite{guss2019minerl} \\ \hline \hline

\end{tabular}
}
\end{table*}

\subsection{Embodied Agents Overview}
\label{sec: paradigm}


Embodied agents are a category of intelligent systems that rely on physical entities to understand, reason, and interact with the physical world \cite{liu2024aligning,duan2022survey,xu2024survey}. 
Embodied agents take advantage of the generalization and reasoning capabilities of LLMs to handle complex tasks and typically comprise two core components: a planner and a controller
\cite{ahn2022can,driess2023palm,song2023llm,brohan2023rt,wang2023voyager,wang2024jarvis,black2024pi_0}.

The \textbf{planner}, often a fine-tuned LLM or MLM, is responsible for comprehending tasks and decomposing them into a series of specific subtasks, known as high-level planning. These subtasks are then passed to the \textbf{controller}, typically a smaller RL model assigned to low-level control. The controller samples actions based on its output \textit{action logits} in each step. 
This paradigm has been widely adopted in recent research \cite{ahn2022can, driess2023palm, song2023llm, brohan2023rt, huang2024embodied, black2024pi_0, wang2024jarvis}, as listed in Table \ref{tab:embodied}. 
\xietong{In parallel, end-to-end vision-language-action (VLA) model is another prevalent approach, which directly generates action tokens or invokes APIs \cite{kim2024openvla, li2023vision, wang2023voyager}. Additionally, many studies focus on models for low-level motion control \cite{brohan2022rt, ghosh2024octo, kalashnikov2021mt, blukis2022persistent, chen2020scanrefer}. Despite these differences, VLAs and motion-control models are functionally analogous to the planner and controller, respectively, and thus align with the same overarching framework.
}
The capabilities of such systems are typically evaluated through standardized simulation benchmarks, such as \cite{shridhar2020alfred, lynch2023interactive, guss2019minerl,o2024open, mees2022calvin, ramakrishnan2021hm3d, liu2023libero}. 

As listed in Table \ref{tab:embodied}, modern embodied agents commonly employ LLM-based planners together with Transformer-based controllers 
for superior multi-task learning capacity \cite{brohan2022rt,lifshitz2023steve, cuhk2023airship, blukis2022persistent, brohan2023rt, figureai2025helix}. However, many such systems \cite{ahn2022can,driess2023palm, song2023llm,wang2023voyager,brohan2023rt} rely on massive (> 50B parameters) planners running on cloud servers, making error-injection studies prohibitively expensive or inaccessible. Moreover, we argue that target tasks should be sufficiently complex and diverse, beyond simple grasping or navigation, to thoroughly assess agent performance. {Therefore}{, in this paper,} we select JARVIS-1 \cite{wang2024jarvis}—a representative open-world agent in Minecraft \cite{guss2019minerl} that meets all these requirements—as our {primary} testbed \xietong{for resilience characterization. 
We further evaluate our method across multiple embodied AI systems \cite{li2023vision, kim2024openvla, brohan2022rt, ghosh2024octo} {to validate the broad applicability}.
}

\begin{figure}[!tb]
    \centering
    \includegraphics[width=1\linewidth]{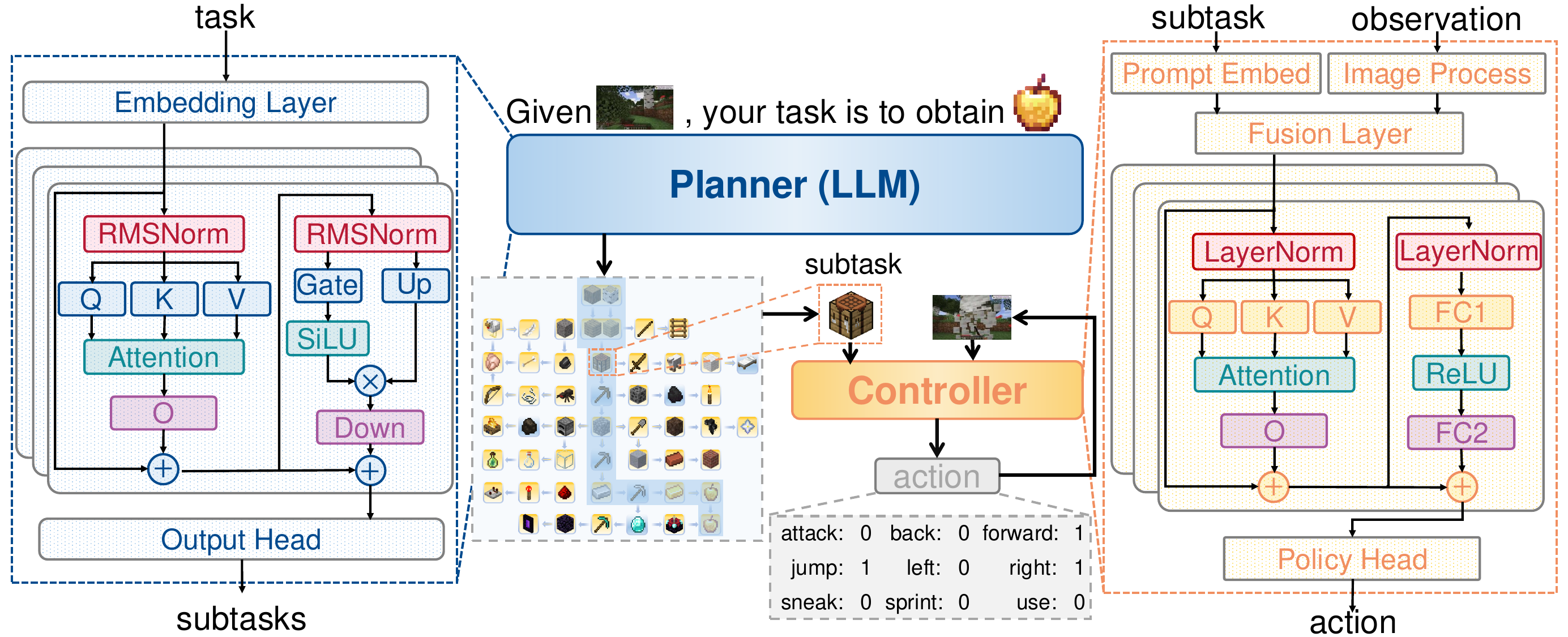}
    \caption{\textbf{JARVIS-1 Platform.} JARVIS-1 consistently acquires high-level items in Minecraft playground. The planner generates a subtask sequence, while the controller integrates subtask prompts with visual observations to determine actions at each step. Both the planner and controller are primarily stacked Transformer blocks, which contain various network components, such as \texttt{K} and \texttt{Down}.
    }
    \vspace{-10pt}
    \label{fig: jarvis}
\end{figure}

Fig. \ref{fig: jarvis} depicts the JARVIS-1 platform,
where the LLM planner and controller, both of which consist of stacked Transformer layers with various \textit{network components} (denoted as \texttt{K}, \texttt{Down}, etc.), work collaboratively within the Minecraft playground. The agent needs to interact with a dynamic 3D environment to acquire tasked items, typically requiring 5$\sim$20 basic subtasks, each involving hundreds of action steps. 
For instance, to obtain an \texttt{enchanted golden apple}, the planner would break it down into subtasks (e.g., \texttt{mine logs}, \texttt{{craft stone pickaxe}}, \texttt{{smelt golden ingots}}) to be executed by the controller in sequence. We refer interested readers to \cite{wang2024jarvis} for more detailed description of the JARVIS-1 platform and Minecraft tasks. Following \cite{wang2024jarvis}, if a subtask remains incomplete after 600 steps, the planner will be re-invoked to adjust the subtask. The high-level task is deemed a failure upon exceeding 12,000 total steps.

In this paper, we evaluate the task quality on multiple Minecraft tasks \cite{wang2024jarvis,wang2023voyager}, ranging from fundamental objectives (e.g., \texttt{wooden pickaxe}, \texttt{stone pickaxe}, etc) to complex ones (e.g., \texttt{iron sword}, \texttt{cooked chicken}, etc), each featuring distinct trajectories and biomes. These tasks are abbreviated as a single teletype word (e.g., \texttt{wooden}) throughout this paper. To mitigate execution randomness, each trial is repeated at least 100 times.
Task quality is measured by success rate ($\uparrow$) and average steps ($\downarrow$), while system efficiency is assessed by average power and total energy consumption. 


\subsection{Computing Platforms of Embodied AI systems}
\label{sec: computing system}


In embodied AI systems, both the planner and controller are typically implemented with DNNs, whose core operations involve dense general matrix multiplications (GEMMs). To meet stringent energy requirements and real-time latency constraints, these DNNs are often deployed on local GPUs \cite{driess2023palm, wen2025tinyvla, brohan2022rt} or on specialized edge accelerators such as TPU-like systolic arrays \cite{jouppi2017datacenter, markidis2018nvidia, brohan2023rt, ahn2022can, huang2025dadu, cuhk2023airship}. In many cases, GEMMs are quantized to low-precision (e.g., INT8) \cite{xiao2023smoothquant, lin2023awq, yuan2023rptq, liu2024spinquant, ashkboos2024quarot, tseng2024quip, kim2025lightrot} to reduce power consumption, particularly in battery-powered systems that must continuously interact with dynamic environments.
In our paper, we consider deploying the entire embodied AI system on a systolic array-based accelerator \xietong{in INT8 format, a standardized and prevalent configuration \cite{zhang2023read,kim2019dris, jouppi2017datacenter, pandey2019greentpu,zhang2018thundervolt, reagen2018ares}}. The planner and controller are orchestrated to execute in a coordinated pipeline. 

\subsection{Fault Tolerance Study}



\textit{\textbf{Error Sources.}}
\xietong{In embodied AI systems, both the memory and logic circuits are susceptible to reliability issues as CMOS technology scales to the nanoscale. Lowering the operating voltage exacerbates this problem: in on-chip SRAMs, reduced voltage noise margins lead to memory bit flips \cite{zhou2010minimizing, wan2024mulberry, reagen2018ares, li2017understanding}, where in logic circuits, voltage underscaling prolongs path delays that would result in 
timing violations and incorrect outputs \cite{zhang2023read, zhang2018thundervolt, jiao2017clim, bal2023novel}. Given that logic circuits typically dominate on-chip power consumption in compute-intensive workloads \cite{casey2009single, tambe2021edgebert, markidis2018nvidia} and that the dynamic nature of embodied AI workloads creates meaningful opportunities for computational optimization, this work focuses on transient computational errors {during inference} induced by voltage underscaling. Extending timing-based resilience to memory components remains promising for future research.}

\textit{\textbf{Model Resilience Study.}}
\label{sec: model resilience study}
\xietong{
Generally speaking, error resilience refers to the ability of an algorithm to tolerate errors, typically quantified by the relationship between BER and model performance \cite{reagen2018ares, wan2021analyzing}.}
Many studies \cite{zhang2023read, sangchoolie2017one, he2020fidelity, hsiao2023silent, papadimitriou2021demystifying, kim2019dris} have adopted random bit-flip models as an abstraction to characterize the inherent resilience of DNNs and autonomous systems. For instance, \cite{li2017understanding} analyzed error propagation in DNNs while \cite{reagen2018ares} explored the relationship between error impacts and data types. \cite{mahmoud2021optimizing} identified heterogeneous vulnerabilities across network components and proposed selective protection. \cite{wan2021analyzing, wan2024mulberry} extended resilience analysis to RL-based applications and autonomous systems. \cite{agarwal2023resilience, xie2025realm} characterized the resilience of individual LLMs and \cite{shah2024characterizing} assessed the risk of 
memory errors in robotic motion planning. However, due to the diversity and collaboration among models, and the complexity of executed tasks, 
the resilience behavior of an embodied agent system may exhibit distinct characteristics.

\textit{\textbf{Error Mitigation Techniques.}}
\label{sec: error mitigation techniques}
Error correction code (ECC) \cite{hamming1950error} can effectively address hardware faults in memory or on-chip buffers. Mitigating transient faults in computational datapaths is more challenging \cite{hsiao2023silent, papadimitriou2021demystifying, wan2021analyzing, bal2023novel}, where various levels of protection can be applied. These methods, though effective in particular scenarios, incur prohibitive hardware or runtime overhead, especially in the context of embodied agents.
For example, at the \textbf{circuit level}, redundancy-based approaches, e.g., dual modular redundancy (DMR) \cite{libano2018selective, khoshavi2020shieldenn,talpes2020compute},  are widely adopted in autonomous vehicles for error detection but require multiple copies of computations to be performed.
Timing-borrowing methods \cite{ernst2003razor, whatmough2018dnn, gundi2020effort} integrate shadow flip-flops (FFs) to detect timing errors in conventional circuits, yet 
lack scalability for large-scale accelerators. \xietong{Although some approaches target accelerators, they either require prior {knowledge of} hardware specification  \cite{zhang2019fault, pandey2019greentpu} or introduce significant hardware {overheads} \cite{zhang2018thundervolt}}, 
{which hinders their applicability for online timing-error mitigation}. 
At the \textbf{algorithm level}, fault-aware retraining, while effective for CNNs \cite{schorn2019efficient, kim2018matic, he2019noise} and simple RL algorithms \cite{wan2021analyzing, wan2024mulberry}, is impractical for embodied agents because of unacceptable retraining costs. 
Lightweight error detection and correction algorithms, such as algorithm-based fault tolerance (ABFT) and its variants \cite{huang1984algorithm, xue2023approxabft, bal2023novel, safarpour2021algorithm}, are suitable for GEMM operations but often require recomputation for error recovery, potentially violating real-time constraints. At the \textbf{system level}, application-specific protection strategies \cite{jha2022exploiting, shah2024characterizing, wan2024vulnerability} for robotics or autonomous systems are tailored to rule-based dataflow and may hardly transfer seamlessly to DNN-driven embodied agents, which may exhibit distinct resilience characteristics.

\vspace{-5pt}
\section{Timing Error Modeling and Injection}
\label{sec: platform}

\subsection{Error Models}
\label{sec: error models}
\begin{figure}[!tb]
    \centering
    \includegraphics[width=\linewidth]{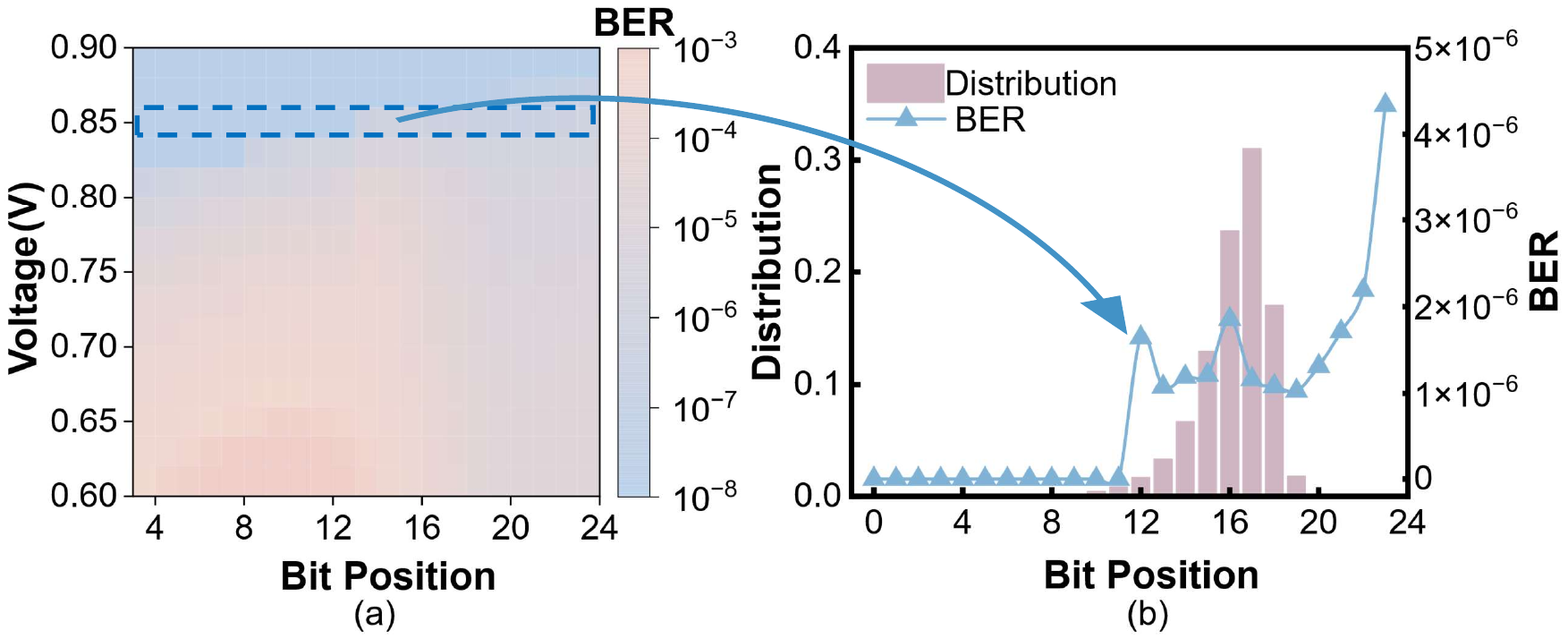}
    \vspace{-20pt}
    \caption{\textbf{Timing error model}. (a) Bit-level timing error rate under different voltages. (b) Error pattern at 0.85V, overlapping with normal runtime activation distribution. 
    }
    \vspace{-10pt}
    \label{fig:error_model}
    
\end{figure}

\vspace{-3pt}
Since memory faults can be effectively mitigated by ECC \cite{mahmoud2021optimizing, hsiao2023silent, papadimitriou2021demystifying, wan2021analyzing, hochschild2021cores}, we focus on transient computational errors during inference, assuming data are correctly fetched from memory.
To model timing errors, we adopt the widely used random bit-flip model \cite{zhang2023read, sangchoolie2017one, he2020fidelity, hsiao2023silent} as an abstraction, with error severity parameterized by BERs. To capture error patterns under different voltages, we synthesize an 8-bit multiplier/24-bit accumulator systolic array with commercial 22nm PDK (nominal voltage: 0.9V) and perform a detailed timing analysis using Synopsys PrimeTime and HSPICE.
The error pattern is shown in Fig.~\ref{fig:error_model}(a), which aligns with prior chip measurements \cite{ernst2003razor, zhang2023read, wan2024mulberry, xie2025realm}. As Fig.~\ref{fig:error_model} illustrates, \textbf{higher bits exhibit frequent large timing errors} that cause severe performance degradation, whereas \textbf{lower bits errors exist within the ordinary runtime data range}. 
\xietong{This correlates with the intuition that higher bits typically have longer carry chains and longer path delays \cite{jiao2017clim, zhang2023read}}.

\vspace{-5pt}

\vspace{-3pt}
\subsection{Error Injection Method}
\label{sec: error injection}
\vspace{-3pt}


We implement a dynamic error-injection framework in PyTorch that emulates bit flips as runtime tensor operations, as commonly used in prior works \cite{sangchoolie2017one, he2020fidelity, hsiao2023silent, papadimitriou2021demystifying,zhang2023read}. 
Following \cite{xiao2023smoothquant}, inputs to GEMM and convolution layers in both the planner and controller are quantized to INT8, and errors are injected into their outputs. For resilience characterization (Sec.~\ref{sec: resilience characterization}), we employ a uniform error model under varying BERs \xietong{to isolate algorithmic resilience from hardware-specific dependencies and derive more generalizable conclusions.} 
In Sec.~\ref{sec: experiment}, we build a look-up table based on Fig.~\ref{fig:error_model}(a) to more accurately capture voltage-induced timing errors.


\begin{figure*}[!tb]
    \centering
    \includegraphics[width=\linewidth]{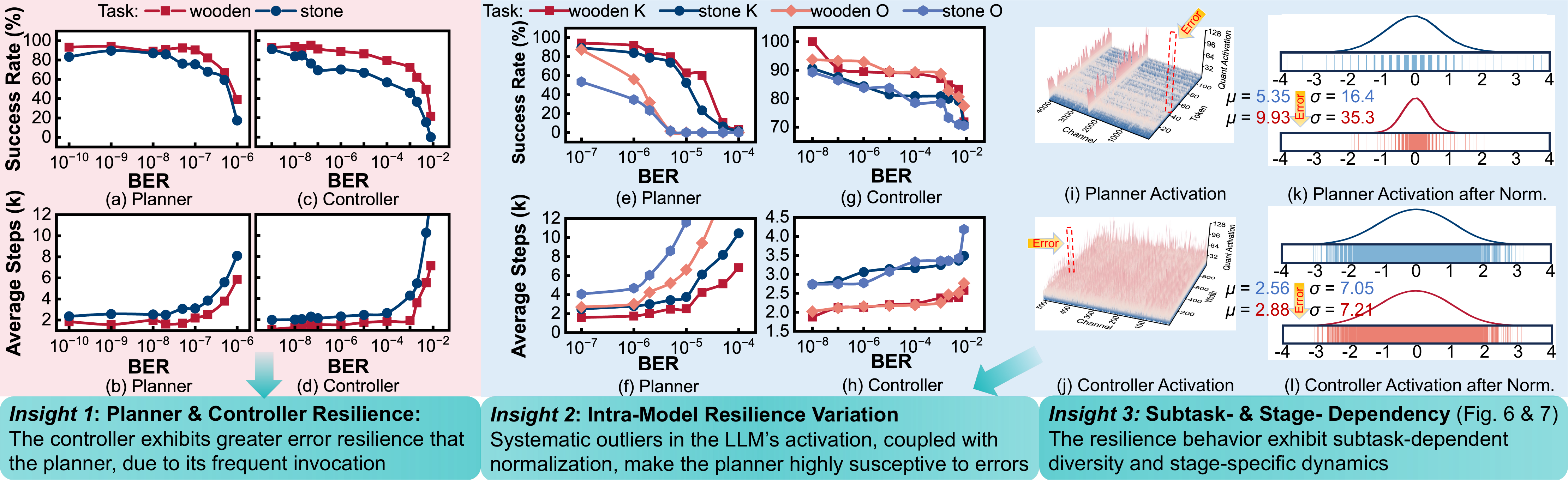}
    \vspace{-10pt}
    \caption{
    \textbf{Resilience Characterization.}
    (a)-(b) Planner resilience characteristics. (c)-(d) Controller resilience characteristics. (e)-(f) Resilience comparison between \texttt{K} and \texttt{O} in the planner. (g)-(h) Resilience comparison between \texttt{K} and \texttt{O} in the controller. (i)-(j) Activation distributions of the pre-norm layer in the planner and controller. (k) Planner normalization outcomes exhibit significant skew under errors; (l) Controller normalization maintains moderate variance.
    }
    \vspace{-5pt}
    \label{fig: resilience characterization}
\end{figure*}

\section{Resilience Characterization}
\label{sec: resilience characterization}


In this section, we present our resilience characterization on embodied AI systems, addressing the following questions:

\textbf{Model-level Q1}:
What are the resilience characteristics of the planner and controller, and how do they differ?

\textbf{Model-level Q2}: 
Furthermore, how does resilience vary among different network components, within the planner and the controller?

\textbf{Application-level Q3}: 
How does resilience behavior dynamically change across subtasks and execution steps?
\vspace{-5pt}
\subsection{Model-level Resilience Characterization}
\label{sec: module level}

\paragraph{Resilience Characteristic of Planner}
\label{sec: planner resilience}
We first inject errors into the planner to evaluate its resilience. Figs. \ref{fig: resilience characterization}(a) and (b) show that success rates plunge near BER = $2\times10^{-8}$ for both tasks, while average steps rise significantly, indicating limited error tolerance. Two factors contribute: (i) 
the planner is invoked only once to guide multiple agent steps, so that a single misleading instruction can lead to prolonged irrelevant or incorrect actions;
and (ii) as an LLM, the planner’s output quality {encounters severe deterioration} under higher BER, producing irrelevant or nonsense text that hinders the controller. 
Further analysis will be provided later.

\vspace{-5pt}
\paragraph{Resilience Characteristic of Controller}
Similarly, we assess the controller’s resilience by tracking the success rate and average steps under varying BER in Figs. \ref{fig: resilience characterization}(c)-(d). A notable drop in success rate and a corresponding rise in average steps emerge around BER = $1\times10^{-4}$, underscoring the controller’s critical impact on task performance yet also revealing notable error tolerance. 
This behavior can be explained from two perspectives as well: (i) the frequent invocation of the controller confines error effects within individual steps and allows correction in subsequent decisions;  (ii) the controller can tolerate suboptimal decisions, as multiple feasible actions may be valid per step. 

\vspace{-7pt}
\paragraph{Discussion} By comparing Figs.\ref{fig: resilience characterization} (a)-(b) with (c)-(d), we conclude that \textbf{the controller exhibits significantly greater error resilience than the planner, owing to its more frequent invocation.}
\vspace{-7pt}


\paragraph{Resilience of different network components}
\label{sec: model level resilience}

This section explores the internal resilience of the planner and controller by injecting errors into individual network components (e.g., \texttt{Q} and \texttt{FC1}) within a Transformer layer. In the LLM planner, the components followed by normalization (including \texttt{O} and \texttt{Down}) 
exhibit noticeably poorer resilience than their counterparts {without normalization} (e.g., \texttt{K}), as shown in Figs.~\ref{fig: resilience characterization}(e)-(f). In contrast, the controller displays only minor variations across components (Figs.~\ref{fig: resilience characterization}(g)-(h)).

This disparity arises from distinct activation distributions. LLMs beyond billions of parameters are extensively reported to produce systematic outliers with much larger magnitudes \cite{xiao2023smoothquant, liu2024spinquant, ashkboos2024quarot} (Fig.~\ref{fig: resilience characterization}(i)), whereas the controller’s activations remain more uniform (Fig.~\ref{fig: resilience characterization}(j)). 
Therefore, in the planner, normalization parameters, i.e., $\mu$ and $\sigma$, 
are subject to these outliers, whereas in the controller, they are more evenly determined across elements.
\xietong{When a single hardware fault occurs and introduces a larger error in LLM computations, these parameters may be drastically skewed 
(e.g., $\mu$ from 5.36 to 9.93 in Fig. \ref{fig: resilience characterization} (k)), 
significantly distorting normalization outcomes. In contrast, the skew in the controller is {relatively} more moderate (e.g., $\sigma$ from 7.05 to 7.21 in Fig. \ref{fig: resilience characterization}(l)), limiting its impact on the outcome.}

\vspace{-7pt}
\paragraph{Discussion} We conclude that \textbf{outliers in the LLM's activation, coupled with subsequent normalization, make the planner highly susceptible to errors.}
\vspace{-7pt}


 

\subsection{Application-level Resilience Characterization}
\label{sec: application level resiliecne}
\vspace{-3pt}

{Embodied agents typically accomplish high-level goals through consecutive sequences of subtasks--often hundreds of steps long.} 
This subsection examines the controller’s resilience across diverse subtasks and execution steps. 

\begin{figure}[!tb]
    \centering
    \includegraphics[width=\linewidth]{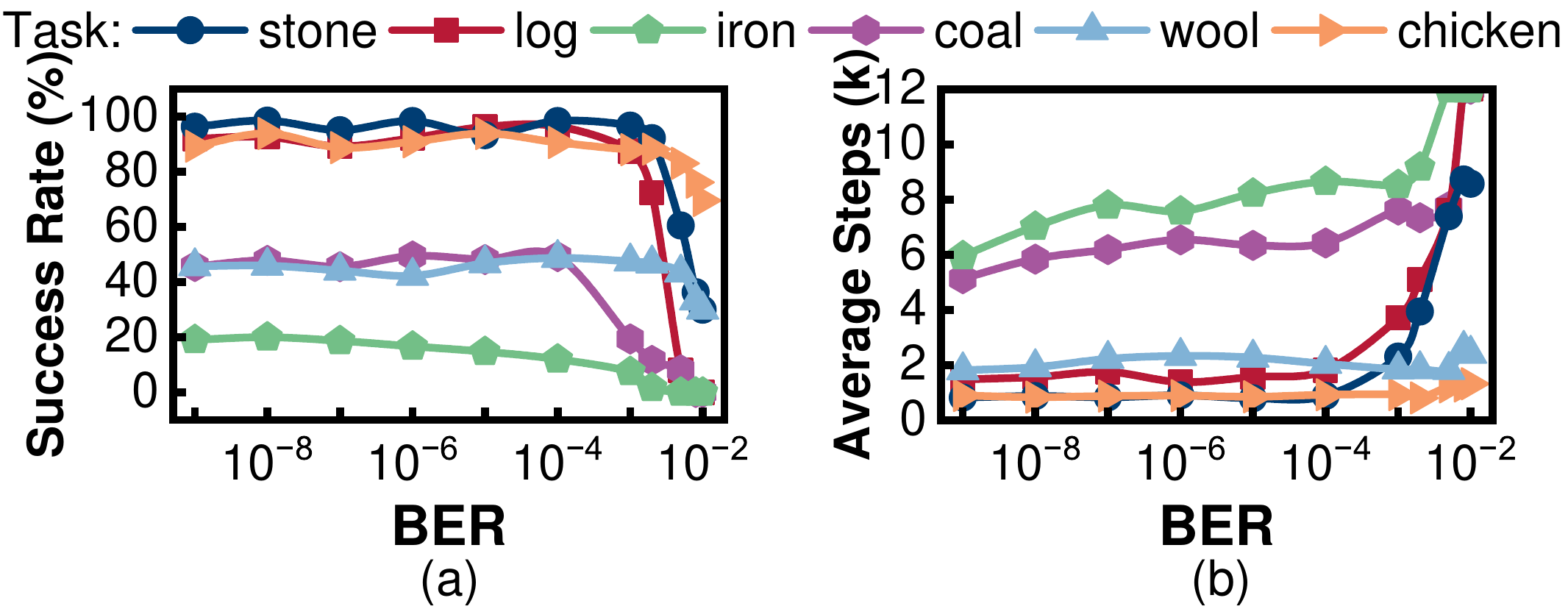}
    \vspace{-20pt}
    \caption{
    \textbf{Different subtasks exhibit diverse resilience. }
    }
    \vspace{-15pt}
    \label{fig: subtask diversity}
\end{figure}

\vspace{-7pt}
\paragraph{Resilience Diversity across Subtasks}
We evaluate resilience across subtasks by injecting errors into controllers performing distinct subtasks in Fig. \ref{fig: subtask diversity}. Results reveal divergent behaviors: subtasks \texttt{log} and \texttt{stone} exhibit abrupt performance degradation beyond BER = $10^{-4}$, while \texttt{chicken} and \texttt{wool} degrade gradually. We attribute this to task structure: sequential subtasks (e.g., tree-chopping) have deterministic action dependencies, where a single error can disrupt entire workflows. In contrast, stochastic tasks (e.g., animal interaction) inherently tolerate variability, alleviating the impact of errors. Thus, error resilience correlates with subtask-specific requirements and environmental randomness.

\begin{figure}[!tb]
    \centering
    \includegraphics[width=1\linewidth]{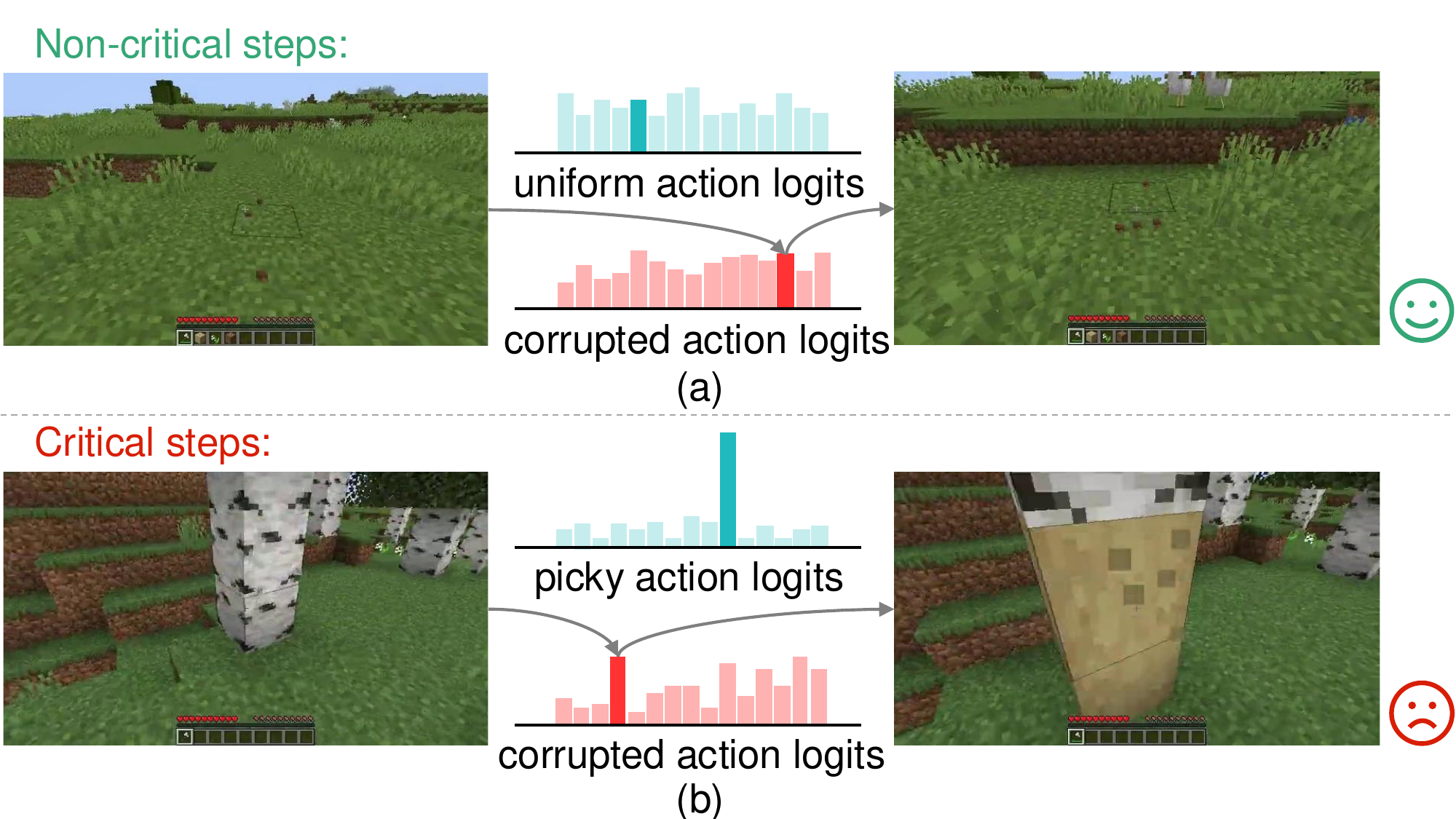}
    \vspace{-20pt}
    \caption{
    \textbf{Stage-specific Resilience.}
    (a) Non-critical steps exhibit uniform action logits, 
    accommodating corruption
    without catastrophic failure. (b) Critical steps exhibit 
    picky
    action logits, where 
    corruptions
    may cause failure (e.g., failing to chop down trees).}
    \label{fig: subtask dynamic}
    \vspace{-10pt}
 
\end{figure}

\vspace{-5pt}
\paragraph{Resilience Dynamics for Execution Steps.}
Embodied agents often transition between critical and non-critical stages within a subtask. To investigate how error impacts vary with execution steps, 
we use the subtask \texttt{mine logs} as a case study and inject bit errors at different steps, as shown in Fig. \ref{fig: subtask dynamic}. 
This subtask can be roughly divided into two phases: exploration (searching for trees) and execution (chopping trees). During exploration, as in Fig. \ref{fig: subtask dynamic}(a), actions are less constrained, and errors primarily cause localized movement deviations without catastrophic failure.
However, the execution stage (Fig. \ref{fig: subtask dynamic}(b)) demands precise, sequential, aligned actions (e.g., striking specific tree blocks). 
where errors can disrupt dependencies, trigger incoherent actions (e.g., misaligned strikes), and eventually prevent task completion.
This demonstrates that error severity 
is stage-specific and increases with task-criticality.

\vspace{-5pt}
\paragraph{Discussion} \textbf{The resilience behavior exhibits subtask-dependent diversity and stage-specific dynamics}.

\section{\method~ Framework}
\label{sec: method}

Building on insights from Sec.~\ref{sec: error models} and \ref{sec: resilience characterization}, we propose \method, a co-optimization \xietong{design principle} encompassing circuit-, model- and application-level techniques for efficient yet reliable embodied AI systems. First, \textit{anomaly detection and clearance} (AD) at the circuit level mitigates the impact of {large errors from} timing violations. Next, \textit{weight-rotation-enhanced planning} (WR) enhances the LLM's resilience against remaining {small but still critical} errors. Finally, an \textit{autonomy-aware voltage scaling} (VS) mechanism monitors the agent’s execution status and dynamically adjusts the voltage of the controller at runtime for additional power savings.
\xietong{Together, by applying AD+WR to the planner and AD+VS to the controller, \method~forms 
a tight vertical integration that delivers
synergistic benefits, enabling aggressive voltage reduction without modifying operating frequency.}

\subsection{Anomaly Detection and Clearance}
\label{sec: AD}

As discussed in Sec. \ref{sec: error models}, lowering the voltage margin often induces timing errors that predominantly flip more significant bits, causing large deviations and performance degradation.
To mitigate this, we propose an \textit{anomaly detection and clearance} (AD) mechanism that identifies out-of-bounds values and clamps them to zero.

\begin{figure}[!tb]
    \centering
    \includegraphics[width=1\linewidth]{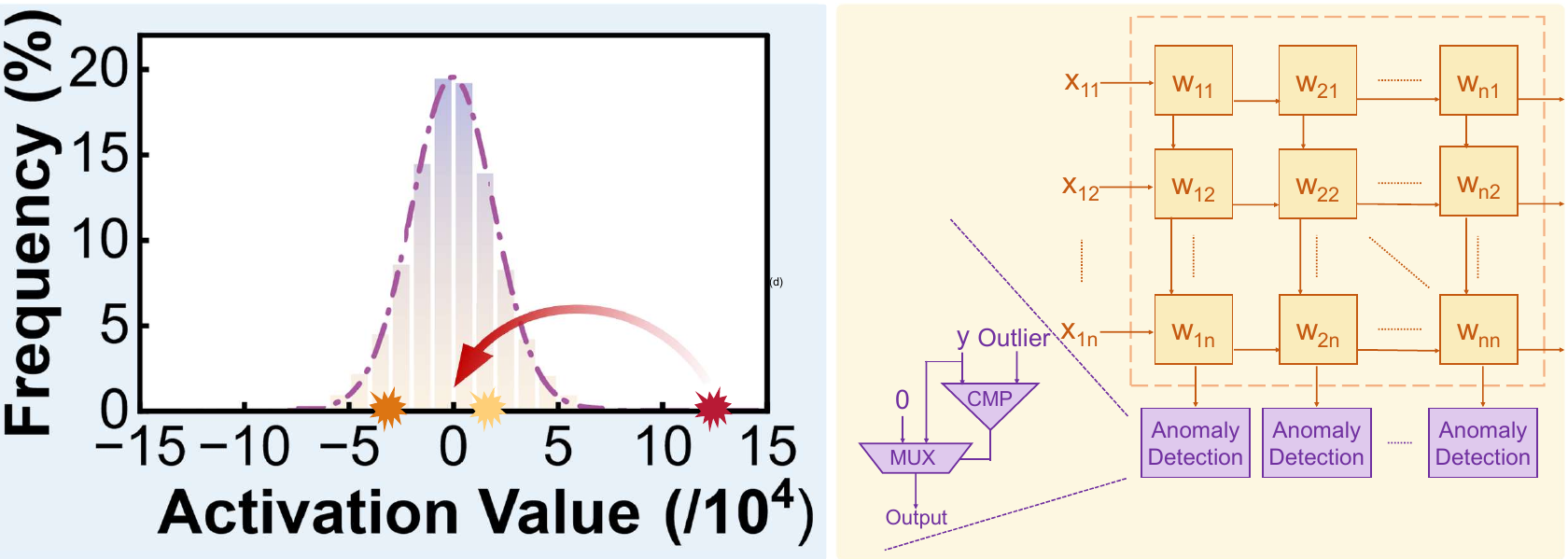}
    \vspace{-12pt}
    \caption{
    \textbf{Anomaly Detection and Clearance.}
    (a) Runtime GEMM results distribution. (b) Circuit design.
    }
    \label{fig: anomaly detection and clearance}
    \vspace{-5pt}
 
\end{figure}

In practice, GEMM operations in accelerators typically use lower-precision (e.g., INT8) formats and re-quantize results based on an offline-determined scaling factor \cite{markidis2018nvidia, jouppi2017datacenter, xiao2023smoothquant}. By profiling runtime GEMM outputs \xietong{throughout the entire JARVIS-1 inference pipeline} (Fig. \ref{fig: anomaly detection and clearance}(a)), we observe two properties: (1) output values rarely occupy the significant bits (as in Fig. \ref{fig:error_model}(b)), and (2) most elements are located near zero. Hence, if any result exceeds the known valid bound (e.g., 127 times the output scaling factor in INT8),
it is flagged as an anomaly and clamped to zero. 
{Although this does not fix faulty results, it substantially alleviates large-magnitude errors, leaving the residual discrepancy to DNNs' inherent fault tolerance.}
\xietong{We emphasize that the two properties mentioned above are well-documented across many DNNs during inference \cite{li2017understanding, yuan2023rptq, ashkboos2024quarot, wan2021analyzing}}. Therefore, our strategy is not restricted to JARVIS-1 but can be generalizable to a wide range of embodied AI systems built upon similar DNN architectures.


We implement AD with a lightweight modification to the systolic array, which supports GEMM operation in embodied AI systems, as shown in Fig. \ref{fig: anomaly detection and clearance}(b). During GEMM computation, weights are stored in each processing element (PE),
while inputs are streamed horizontally from left to right. Partial sums accumulate along columns as they propagate vertically downward, with the final results generated at the bottom. We augment this structure with a row of \textit{anomaly detection units} at the output stage. Each unit consists of a comparator and a multiplexer, and the final results are checked against the valid range. Any out-of-range anomaly is clamped to zero, while in-range values pass through unchanged. 
\xietong{This post-processing detection decouples from hardware specification and incurs minimal overhead, ensuring portable and lightweight real-time error suppression.}

\subsection{Weight-Rotation-Enhanced Planning}
\label{sec: Rotation}

As discussed in Sec. \ref{sec: model level resilience}, pre-normalization components in the planner LLM are highly vulnerable to errors. While AD in Sec.~\ref{sec: AD} suppresses timing errors exceeding the anomaly bound, it does not address smaller errors within the normal range that still degrade task quality. To remedy this, we propose \textit{weight rotation-enhanced planning} (WR), a model-level approach that redistributes activation outliers via offline weight transformations.

Inspired by quantization techniques \cite{liu2024spinquant, ashkboos2024quarot, tseng2024quip, kim2025lightrot}, we \textit{rotate} activations into an outlier-free distribution via multiplication with a rotation matrix. 
A simple yet effective choice is the standard Hadamard matrix $\mathbf H$ \cite{hedayat1978hadamard}, which is recursively defined via the Kronecker product ($\otimes$) \cite{van2000ubiquitous}:
$$
\mathbf{H}_2=\frac{1}{\sqrt{2}}\begin{bmatrix}1&1\\1&-1\end{bmatrix}\quad\mathrm{~and~}\quad\mathbf{H}_{2^k}=\mathbf{H}_2\otimes\mathbf{H}_{2^{k-1}}
$$
This rotation disperses outliers across dimensions, yielding more uniform activations. 
Crucially, since the rotation matrices can preserve the L2 norm as RMSNorm denominators, they integrate seamlessly into LLM computation without altering overall network outputs.



\begin{figure}[!tb]
    \centering
    \includegraphics[width=1\linewidth]{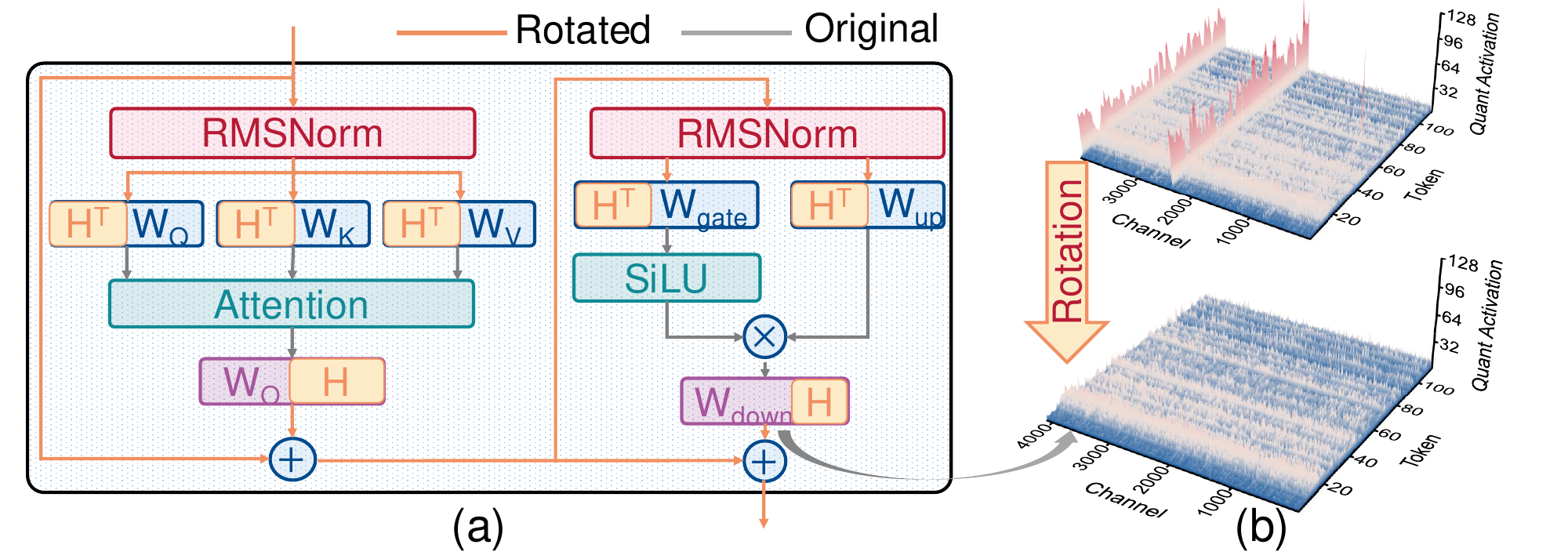}
    \vspace{-12pt}
    \caption{ \textbf{Weight-Rotation Enhanced Planning.} (a) Weights are multiplied offline by Hadamard matrices. (b) Pre- and post-rotation activation distribution. 
    }
    \label{fig: weight enhanced rotation planning}
    \vspace{-5pt}
 \end{figure}

\xietong{Our technique focuses solely on redistributing activations before normalization and avoids online rotations as in \cite{liu2024spinquant,ashkboos2024quarot, kim2025lightrot, tseng2024quip}. Fig. \ref{fig: weight enhanced rotation planning}(a) depicts the required rotations, which can be merged into the weights (\texttt{O} and \texttt{Down}) offline.}
For example, consider \texttt{O} with input $\mathbf X$ and weight $\mathbf W_{\rm O}$, yielding $\mathbf Y=\mathbf X \mathbf W_{\rm O} $. By right-multiplying $\mathbf W_{\rm O}$ with $\mathbf H$, we obtain $\mathbf X (\mathbf W_{\rm O}\mathbf H )=(\mathbf X \mathbf W_{\rm O})\mathbf H = \mathbf{YH}$ as an outlier-free activation in Fig. \ref{fig: weight enhanced rotation planning}(b). To maintain the functional correctness of SiLU in the multi-layer perceptron (MLP) block, $\mathbf{W}_{\rm gate}$ and $\mathbf{W}_{\rm up}$ are left-multiplied by $\mathbf{H}^{\rm T}$.
Similarly, to eliminate outliers after \texttt{Down}, $\mathbf W_{\rm down}$ are right-multiplied by $\mathbf{H}$, while 
$\mathbf W_{\rm Q}$, $\mathbf W_{\rm K}$ and $\mathbf W_{\rm V}$ in the next layer are left-multiplied by $\mathbf{H}^{\rm T}$. 
These offline procedures introduce no additional runtime overhead 
\xietong{and require no model-specific finetuning}.
Additionally, rotated activations can collaborate with AD by further tightening the anomaly bound, enabling more effective error suppression.

\subsection{Autonomy Adaptive Voltage Scaling}
\vspace{-3pt}
\label{sec: Entropy}


To leverage the subtask- and step-dependent resilience behavior (Sec. \ref{sec: application level resiliecne}), we propose an \textit{autonomous-adaptive voltage scaling} (VS) mechanism to maximize controller efficiency. We first identify the entropy \cite{ash2012information} of action logits as a principal runtime status indicator and train a lightweight network to predict it before inference. We then introduce an LDO-based architecture for dynamic voltage adjustment {with minimal circuit overhead}, offering a generalizable runtime voltage scaling solution for RL-based embodied controllers.

\vspace{-5pt}
\paragraph{Entropy of Action Logits as Indicator}
Embodied agents typically face dynamic environments, where each step’s importance varies remarkably. As in Sec. \ref{sec: application level resiliecne}, resilience behavior differs across subtasks and dynamic execution statuses, making manually designed or fixed policies inadequate.

\begin{figure}[!tb]
    \centering
    \includegraphics[width=1\linewidth]{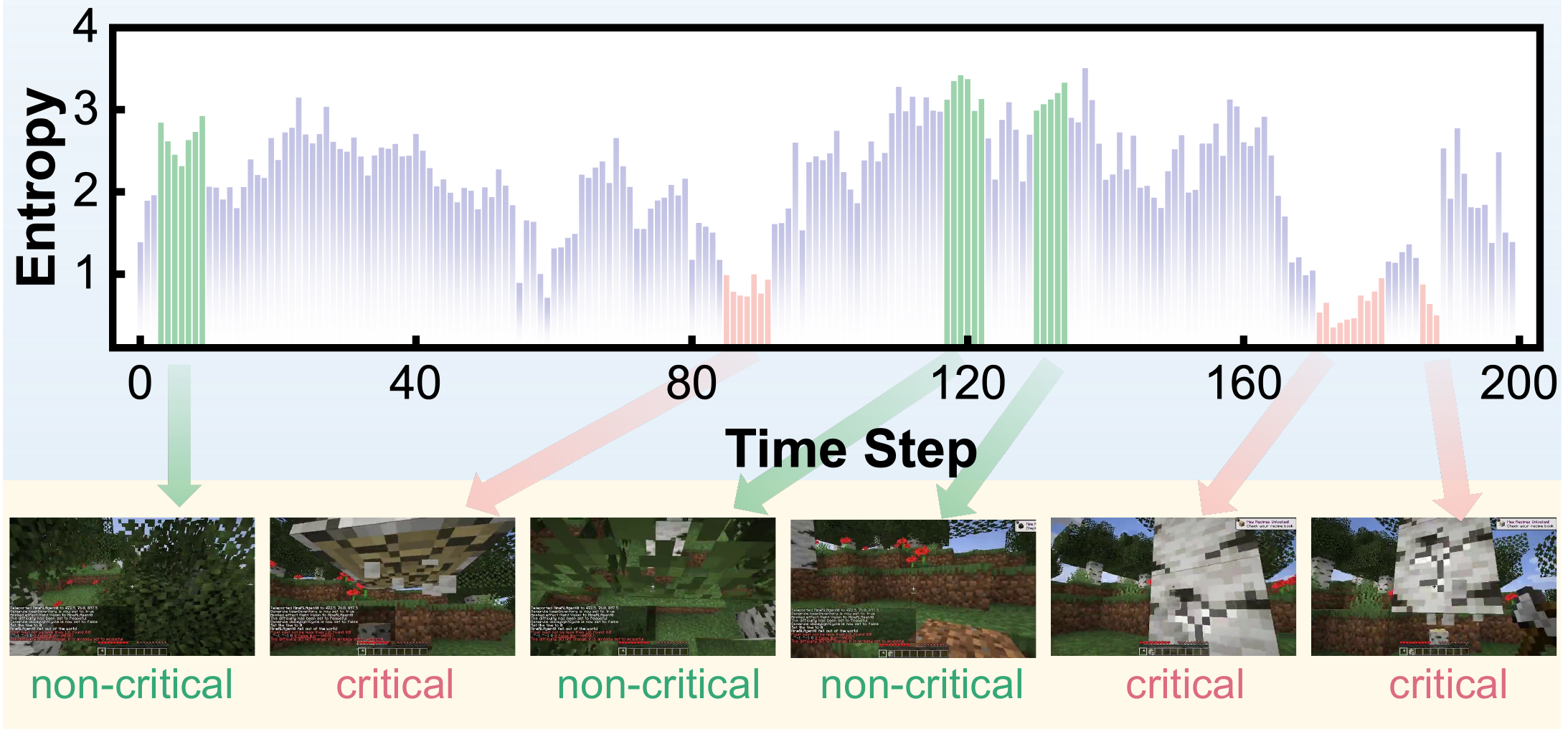}
    \caption{
    \textbf{Entropy curve across timesteps.} Higher entropy corresponds to non-critical execution steps, while lower entropy indicates critical ones.}
    \label{fig: entropy}
    \vspace{-10pt}
 
\end{figure}

To design an effective voltage-control policy, a runtime metric is required
to quantify step-criticality. Observing the action logits (denoted as $\boldsymbol z\in \mathbb R^n$) in Fig. \ref{fig: subtask dynamic} reveals stage-dependent distributions: near-uniform logits during non-critical phases indicate low confidence, while sharply picky logits in critical phases reflect high confidence in specific actions. This motivates us to use the entropy \cite{ash2012information}, defined as
$
H(\boldsymbol z)=-\sum\limits_{i=1}^{n}z_i\log(z_i)
$
, to quantify this confidence and represent step criticality. Fig. \ref{fig: entropy} shows an example of entropy across task steps, which precisely indicates the agent’s status. Lower
entropy corresponds to critical stages that require robust voltage margins, whereas higher entropy indicates non-critical phases amenable to lower voltages for better efficiency. This approach dynamically aligns algorithmic certainty with system energy costs.

\begin{figure}[!tb]
    \centering
    \includegraphics[width=1\linewidth]{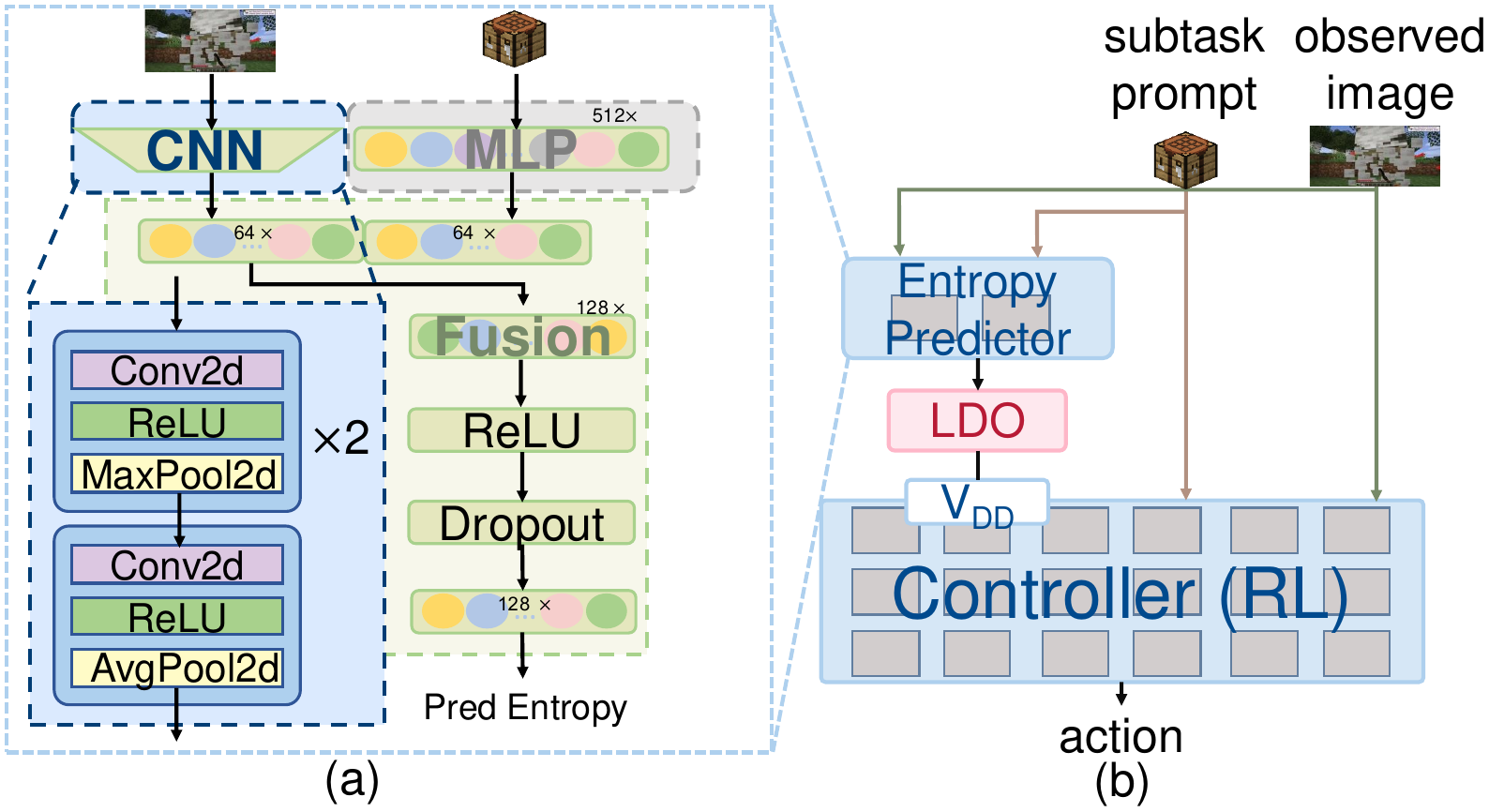}
    \vspace{-15pt}
    \caption{
    \textbf{Autonomy-Aware Voltage Scaling.}
    (a) Entropy predictor network architecture. (b) Voltage scaling system: an LDO dynamically adjusts the controller model’s supply voltage based on real-time entropy predictions.}
    \label{fig: DVFS system}
    \vspace{-10pt}
 
\end{figure}

\vspace{-5pt}
\paragraph{Entropy Predictor}
To enable step-granularity voltage scaling, we introduce a pre-execution entropy predictor.
Taking in the subtask prompt embedding and observed image, the entropy predictor estimates the controller's error-free entropy that avoids distortion by any prior errors.
As depicted in Fig. \ref{fig: DVFS system}(a), the predictor uses a CNN for image processing, an MLP for prompt embeddings, and a fusion MLP to output a single scalar entropy estimate.
\xietong{The model details are provided in Appendix \ref{appendix: model}}. 
We construct a dataset of over 250,000 frames from diverse tasks, each containing a prompt embedding, an observed image, and a ground-truth entropy value derived from error-free controller executions. The predictor is trained with a mean-square error (MSE) loss between the predicted and ground-truth entropy.


\vspace{-5pt}
\paragraph{Voltage Scaling System}
Our voltage scaling 
system adaptively adjusts the supply voltages of the controller model based on the predicted entropy, as shown in Fig.~\ref{fig: DVFS system}(b). The controller model and predictor all run on PE arrays, while the predictor operates at nominal voltage to ensure an error-free entropy prediction. This prediction
is passed to digital LDO \cite{tambe2021edgebert}, which adjusts the controller's operating voltage.
Higher entropy triggers lower voltages, with the exact entropy-to-voltage mapping determined in Sec. \ref{sec: DVFS evaluation}. The controller’s voltage is updated every 5 steps—sufficiently frequent to track workload changes yet still sparse enough to minimize switching overhead (Sec. \ref{sec: update interval}).

\section{Experiments}
\label{sec: experiment}

This section first introduces our experiment setup (Sec.~\ref {sec: experiment setup}), and circuit-level metrics (Sec. \ref{sec: circuit evaluation}). 
We then analyze the effectiveness of the three core techniques of \method~individually (Sec.~\ref {sec: AD evaluation}–\ref{sec: DVFS evaluation}) 
and their interactions via ablation studies (Sec.~\ref{sec: ablation}). 
We further demonstrate the overall benefits of \method, validating its generality across diverse tasks and \xietong{embodied AI platforms (Sec.~\ref{sec: multi-task}), and analyzing its real-world impact (Sec. \ref{sec: energy}). 
Finally, we discuss the sensitivity to several configurations (Sec. \ref{sec: configuration})} and compare our proposed approach against existing methods (Sec.~\ref{sec: existing techniques}).

\subsection{Experiment Setup}
\label{sec: experiment setup}

\begin{figure}[!tb]
    \centering
    \includegraphics[width=1\linewidth]{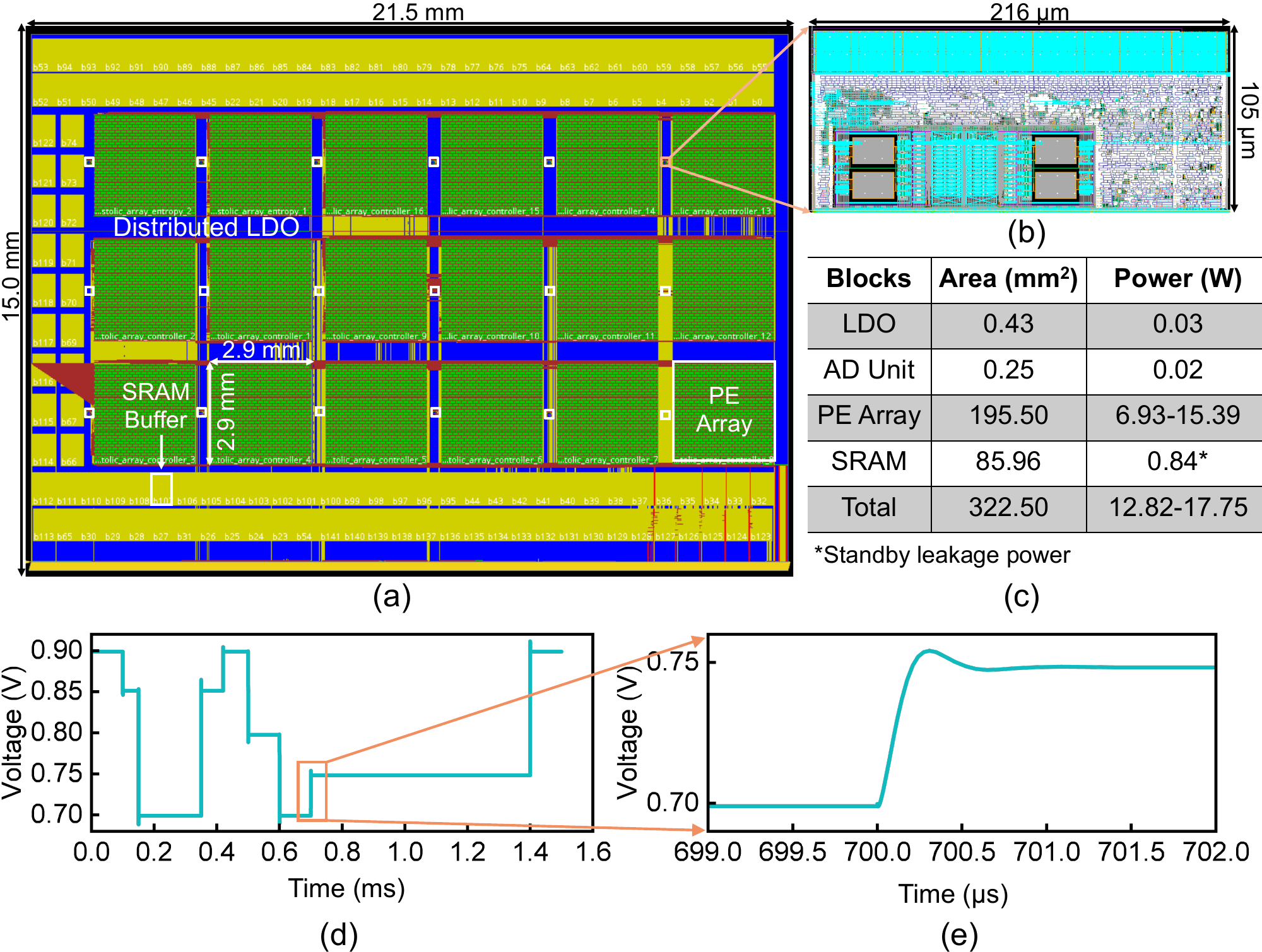}
    \vspace{-20pt}
    \caption{
    \textbf{Hardware Platform.}
    (a) Post-PnR layout of the full accelerator. (b) Detailed layout of LDO. (c) Area and power breakdown. (d)(e) Example voltage scaling waveforms.}
    \label{fig: layout} 
\end{figure}

\begin{figure*}[!tb]
    \centering
    \includegraphics[width=1\linewidth]{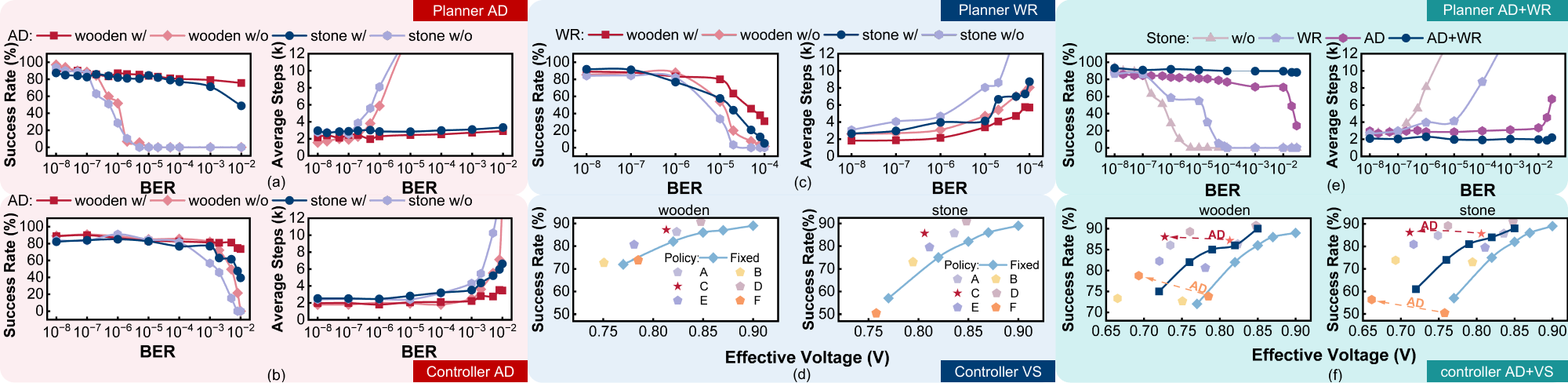}
    \vspace{-8pt}
    \caption{
    \textbf{Evaluation of CREATE Techniques.}
    (a) AD evaluation on the planner. (b) AD evaluation on the controller. (c) WR evaluation on the planner. (d) VS evaluation on the controller. (e) Ablation study (AD+WR) on the planner. (f) Ablation study (AD+VS) on the controller. }
    \label{fig: techniques result}
    \vspace{-5pt}
 
\end{figure*}

\textit{\textbf{Hardware Platform.}}
We integrate \method~ techniques in a unified hardware system designed for end-to-end embodied AI deployment, as shown in Fig. \ref{fig: layout}. Anomaly detection units in Sec. \ref{sec: AD} are appended to 
systolic arrays with $128\times 128$ PEs that are composed of an 8-bit multiplier and a 24-bit accumulator. The dynamic voltage scaling system required in Sec. \ref{sec: Entropy} is supported with a distributed LDO design. Our design is synthesized with a commercial 22 nm PDK at a nominal 0.9 V supply voltage and a 2 ns clock cycle. The LDO is built based on \cite{kim20210} and can dynamically scale the PE array voltage from 0.6 V to 0.9 V in 10 mV steps. 
The LDO achieves 99.8\% peak current efficiency and a 90 ns/50 mV transient response time, as listed in Table \ref{tab:LDO}. 
\xietong{On-chip SRAM buffers (142 $\times$ 512 KB, totaling 71 MB) are placed around the PE arrays, with HBM2 \cite{jun2017hbm} used for off-chip memory.
During inference, weights and inputs are initially loaded from off-chip HBM2 and cached in on-chip SRAM buffers for reuse, with activations also stored on-chip. For the controller, given the relatively small model size, our SRAM capacity can fully accommodate all weights, thereby preventing repeated off-chip memory accesses per invocation.}
Figs. \ref{fig: layout}(a) and (b) show the Post-PnR layout of the full system and an individual LDO, respectively,  with example voltage-scaling waveforms in Figs. \ref{fig: layout}(d) and (e). The area and power of primary blocks in Fig. \ref{fig: layout}(c) are reported based on post-layout simulations. Cycle-level behaviors, \xietong{including inference latency and memory access, are modeled based on SCALE-Sim \cite{samajdar2018scale}.}


\begin{table}[!tb]
    \centering
    \caption{Performance specifications of the LDO.}
    \vspace{-5pt}
    \label{tab:LDO}
    \renewcommand*{\arraystretch}{1.05}
    \resizebox{\linewidth}{!}{
    \begin{tabular}{ lr | lr } 
        \hline\hline
         Technology &  22\,{nm} &   Area & 0.43\,$\rm mm^2$  \\      
         $V_{\mathrm{out}}$ & 0.6--0.9\,{V} & $I_{\mathrm{load,max}}$ & 15.2\,{A} \\ 
         $t_{\rm resp}$ &90\,ns/50\,mV&$\eta_{\rm peak}$ & 99.8\% @ $I_{\mathrm{load,max}}$ \\  
         $V_{\rm step}$&10 mV& $J$& 35 $\rm A/mm^2$ \\
         \hline\hline  
         \multicolumn{4}{@{}l@{}}{\footnotesize $t_{\mathrm{resp}}$: response time, $\eta_{\mathrm{peak}}$: peak current efficiency, $J$: current density.} \\
    \end{tabular}
    \vspace{-40pt}
    }
\end{table}


\textit{\textbf{Entropy Predictor Training.}}
We evaluate the proposed entropy predictor by integrating it into real-time mission episodes. 
In our system, the theoretical maximum entropy of action logits is 13.07, with the majority of values less than 4.
For training details, we employ an MSE loss between the predicted and ground-truth entropy values, using AdamW optimizer \cite{loshchilov2019decoupledweightdecayregularization} with a weight decay of $1\times 10^{-2}$. The model is trained for 200 epochs with a batch size of 128 and an initial learning rate of $1\times 10^{-4}$.

\textit{\textbf{Evaluation Tasks.}}
We first evaluate our framework using the fundamental \texttt{wooden} and \texttt{stone} tasks as primary benchmarks in Sec. \ref{sec: AD evaluation}-\ref{sec: ablation}, presenting its performance and ablation outcomes for rapid evaluation. Each experiment is repeated over 100 times to reduce randomness. 
To validate generality, we further test our framework on six unseen and more challenging benchmarks in JARVIS-1, which feature distinct task trajectories, 
providing evidence for its robustness and adaptability.
\xietong{
To evaluate cross-platform generality, we extend our approach to separate platforms \cite{kim2024openvla, brohan2022rt,ghosh2024octo, li2023vision} for the planner and controller. 
For the planner, we transplant our implementation to OpenVLA \cite{kim2024openvla} on LIBERO \cite{liu2023libero} and to RoboFlamingo \cite{li2023vision} on  CALVIN \cite{mees2022calvin}. For the controller, we integrate our approach with Octo \cite{ghosh2024octo} and RT-1 \cite{brohan2022rt} on different OXE \cite{o2024open} tasks.}
Timing errors are simulated by randomly flipping bits in the accumulation results following \cite{jiao2017clim, zhang2023read, sangchoolie2017one,he2020fidelity,hsiao2023silent}. The relation between BER and operating voltage follows Fig. \ref{fig:error_model}(a).

\textit{\textbf{Evaluation Metrics.}}
We evaluate both task performance (success rate, average steps) and system efficiency (average power, total energy).
Success rate measures the percentage of completed trials, while average steps denote the mean steps among all successful trials.
To briefly measure the power consumption, we introduce \textit{effective voltage}, defined as the constant voltage yielding equivalent total energy {consumption} to actual varying voltages.
The computational energy is simulated by tracking per-step operating voltages for planner, controller, and entropy predictor execution.  The switching power is negligible in practice. 
Failed tasks are included in energy calculations by assuming full execution.

\subsection{Circuit Overhead Comparison}
\label{sec: circuit evaluation}
Fig. \ref{fig: layout}(b) presents the area and power breakdown of the proposed system. The anomaly detection units integrated into PE arrays introduce 0.08\% area and 0.10\% power overhead, while the LDOs distributed between arrays contribute 0.13\% area and 0.14\% power overhead—both negligible.

As summarized in Table \ref{tab:latency}, the full system achieves inference latencies of 11.2 ms for the planner, 942 $\rm \upmu s$ for the controller, and 8.57 $\rm \upmu s$ for the entropy predictor, respectively. Voltage switching latency remains bounded below 540 ns, which is orders of magnitude lower than the controller’s inference latency. These meet our embodied AI system's 30 Hz real-time inference requirement \cite{wang2024jarvis}.

\begin{table}[!tb]
    \centering
    \caption{Performance of the full accelerator. 
    }
    \vspace{-3pt}
    \label{tab:latency}   
    \renewcommand*{\arraystretch}{1.08}
    \resizebox{\linewidth}{!}{
    \begin{tabular}{ c c| cc } 
        \hline\hline
        Peak Performance &  144 TOPS & Switching Latency & 540 ns \\
        Planner MACs &5.3 T&Planner Latency & 11.2 ms \\
        Controller MACs &102 G& Controller Latency & 942 $\upmu \rm s $ \\
        Predictor MACs &43 M& Predictor Latency & 8.57 $\upmu \rm s $ \\     
        \hline\hline  
        \multicolumn{4}{@{}l@{}}{\footnotesize TOPS: tera operations per second. MACs: multiplication-and-accumulations. } 
        \vspace{-3pt}
    \end{tabular}
    }
\end{table}

\subsection{Anomaly Detection and Clearance Evaluation}
\label{sec: AD evaluation}



 

\textit{\textbf{Planner Performance Comparison.}}
To evaluate the effectiveness of AD on the planner, Fig. \ref{fig: techniques result}(a) compares success rates and average steps with and without AD. AD significantly improves robustness against timing errors: at BER = $1\times10^{-5}$, success rates increase from 0\% to 85\% for \texttt{wooden} and from 0\% to 83\% for \texttt{stone}, respectively, \xietong{while average steps fall back to error-free scenarios}. With negligible circuit overhead, AD sustains comparable performance at BER$<1\times 10^{-5}$, unlocking substantial potential energy savings.

\textit{\textbf{Controller Performance Comparison.}}
The impact of AD on the controller is evaluated in Fig. \ref{fig: techniques result}(b). AD significantly enhances the controller’s robustness: success rates improve by up to 73 and 48 percentage points for \texttt{wooden} and \texttt{stone}, while average steps are reduced by 51\% and 49\% at BER = $5\times 10^{-3}$, respectively. AD achieves particularly notable controller performance gains at BER > $1\times 10^{-4}$, revealing its effectiveness in mitigating timing errors under aggressive voltage scaling.

\subsection{Weight Rotation Enhanced Planning Evaluation}
\label{sec: Rot Evaluation}

Fig. \ref{fig: techniques result}(c) evaluates the impact of WR on planner reliability. By redistributing activation distributions in the LLM planner, this method prevents an individual error from drastically skewing normalization outcomes. WR improves success rates by 43\% (\texttt{wooden}) and 40\% (\texttt{stone}) while reducing average steps by 33\% and 49\% at BER = $2\times 10^{-5}$, respectively, which underscores the efficacy of WR in strengthening the planner's robustness under timing errors.

\subsection{Autonomy Adaptive Voltage Scaling Evaluation}
\label{sec: DVFS evaluation}

\begin{figure}[!tb]
    \centering
    \includegraphics[width=1\linewidth]{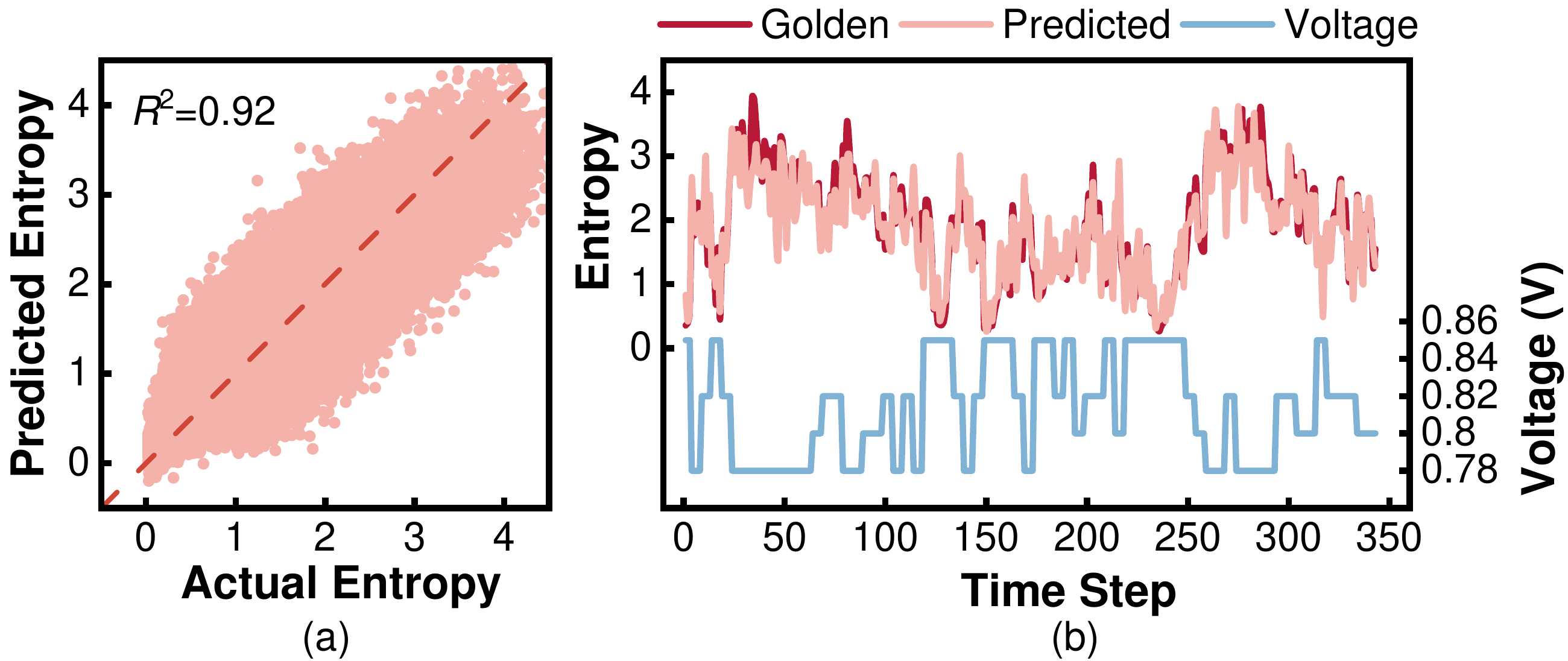}
    \vspace{-18pt}
    \caption{\textbf{Entropy predictor accuracy evaluation.} (a) The predicted entropy and actual entropy are strongly correlated, with $R^2=0.92$. (b) Real-time entropy prediction and the corresponding operating voltage level. Entropy predictor can effectively guide voltage scaling.}
    \label{fig: entropy predictor accuracy}
    \vspace{-10pt}
\end{figure}

\noindent \textit{\textbf{Entropy Predictor Accuracy and Analysis.}}
Our training process converges to a final MSE of $9.96\times10^{-2}$. Fig. \ref{fig: entropy predictor accuracy}(a) demonstrates a strong correlation ($R^2=0.92$) between predicted and ground-truth entropy. 
Fig. \ref{fig: entropy predictor accuracy}(b) depicts the runtime performance, where
the predicted entropy closely tracks actual entropy dynamics across time steps, assuring its utility for real-time voltage scaling.

\textit{\textbf{Performance Improvements.}}
To demonstrate the impact of VS, we compare our method against a constant-voltage baseline that maintains a fixed voltage throughout execution. Task performance is measured by the success rate on the \texttt{wooden} and \texttt{stone} benchmarks, while energy efficiency is quantified by the effective voltage, as shown in Fig.~\ref{fig: techniques result}(d). \xietong{This figure includes six entropy-to-voltage mapping policies A–F (Fig. \ref{fig:entropy-to-voltage}) identified through a search over 100 candidates, each offering a superior balance between performance and efficiency. Compared to constant-voltage policies (blue line), which exhibit declining success rates at lower voltages, our adaptive policies consistently achieve higher energy efficiency and task performance. Among them, Policy C (red star) reduces effective voltage by 7.3\% without sacrificing success rates. Therefore, it is chosen as the optimal policy, which advances the reliability–efficiency Pareto frontier and serves as the default configuration in subsequent experiments.
}


 
\begin{figure}[!tb]
    \centering
    \includegraphics[width=1\linewidth]{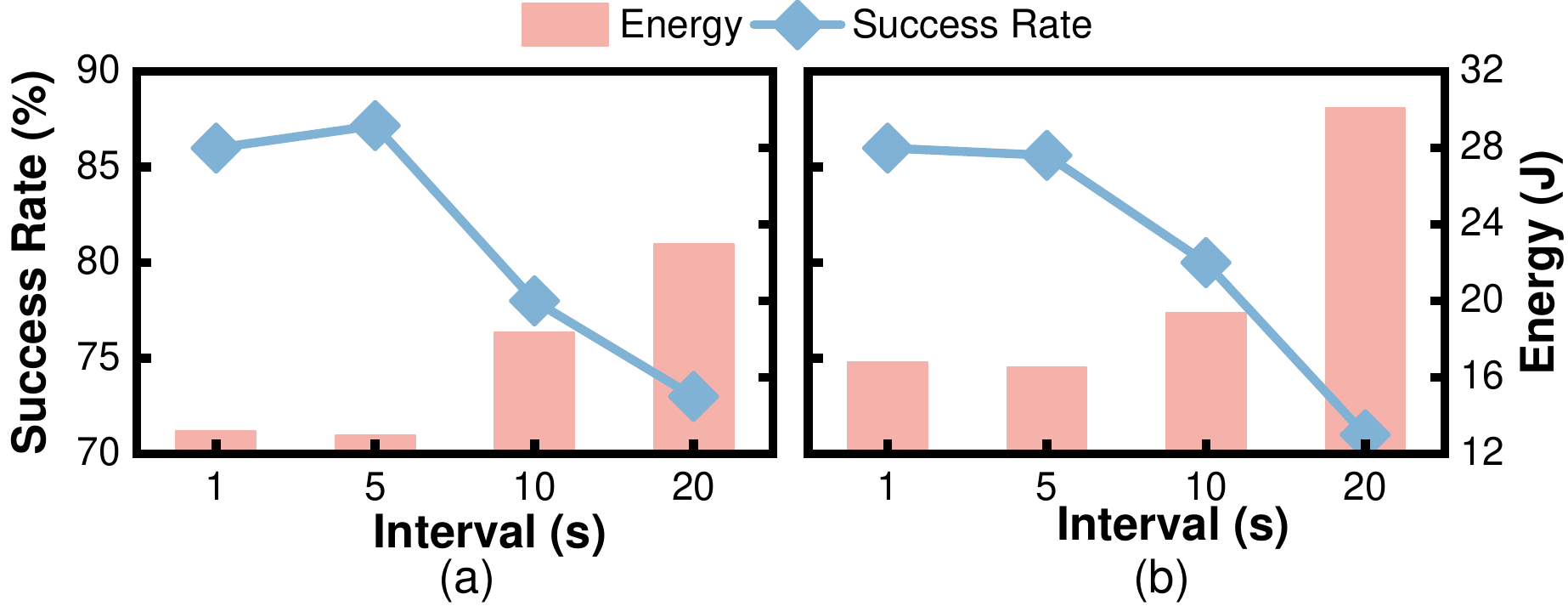}
    \caption{
    \textbf{Voltage update interval effects} on energy consumption and success rate for (a) \texttt{wooden} and (b) \texttt{stone}.}
    \label{fig: voltage update interval}
    \vspace{-10pt}
 
\end{figure}

\textit{\textbf{Voltage Update Interval Analysis.}}
\label{sec: update interval}
This paragraph evaluates the impact of the voltage update interval, i.e., the number of steps between voltage adjustments. As shown in Fig.~\ref{fig: voltage update interval}, intervals of 1 and 5 steps sustain high success rates for both tasks, whereas intervals of 10 and 20 steps degrade performance, indicating inadequate real-time response. Energy analysis further reveals that a 5-step interval incurs slightly lower overhead than a 1-step interval. Therefore, a 5-step interval strikes an optimal balance between timely adaptation and VS efficiency and is thus chosen as the default voltage update interval.

\vspace{-5pt}
\subsection{Ablation Study}
\vspace{-3pt}
\label{sec: ablation}

We evaluate individual techniques in our framework via ablation studies. As AD and WR target the planner while AD and VS optimize the controller, we isolate their contributions by testing AD+WR and AD+VS configurations separately.

\textit{\textbf{Ablation Study on Planner.}}
As shown in Fig. \ref{fig: techniques result}(e), while both AD and WR can independently enhance robustness, their combination yields significantly stronger resilience, preserving task quality even at BER=$1\times 10^{-2}$.
This not only suggests the efficacy of AD and WR separately, but also showcases a potential synergistic effect.
We attribute this synergy to WR’s ability to distribute outliers across dimensions, which lowers AD's anomaly detection threshold and further mitigates error effects.

\textit{\textbf{Ablation Study on Controller.}}
\xietong{Fig.~\ref{fig: techniques result}(f) compares VS combined with AD (darker lines and pentagons) against standalone VS (lighter lines and hexagons). By improving robustness at low voltages, AD enables VS to operate at further reduced voltage levels, bringing in more energy efficiency.
This effect is visualized by the leftward shifts in the data points, with dashed arrows indicating reductions in effective voltage. When collaborating with AD, the optimal policy lowers effective voltage by 11.2\% (\texttt{wooden}) and 13.2\% (\texttt{stone}) relative to constant-voltage baselines with AD, and by 16.4\% and 18.2\% versus baselines without AD, all while maintaining task success rates. These results highlight another synergistic benefit of combining VS and AD.

}

\subsection{Overall Evaluation across Different Tasks}
\label{sec: multi-task}

This section evaluates the overall benefits of the entire framework.
To further evaluate the generality, we extend experiments to \xietong{six additional, more complex workloads within JARVIS-1, including \texttt{charcoal},  \texttt{chicken}, \texttt{coal}, \texttt{iron}, \texttt{wool}, and \texttt{seed} tasks}. We first report success rates and \xietong{computational} energy consumption at 0.75V to demonstrate reliability improvement for low voltage operation. We then quantify efficiency gains through energy savings when operating under the lowest voltage that sustains an acceptable performance.

\begin{figure}[!tb]
    \centering
    \includegraphics[width=1\linewidth]{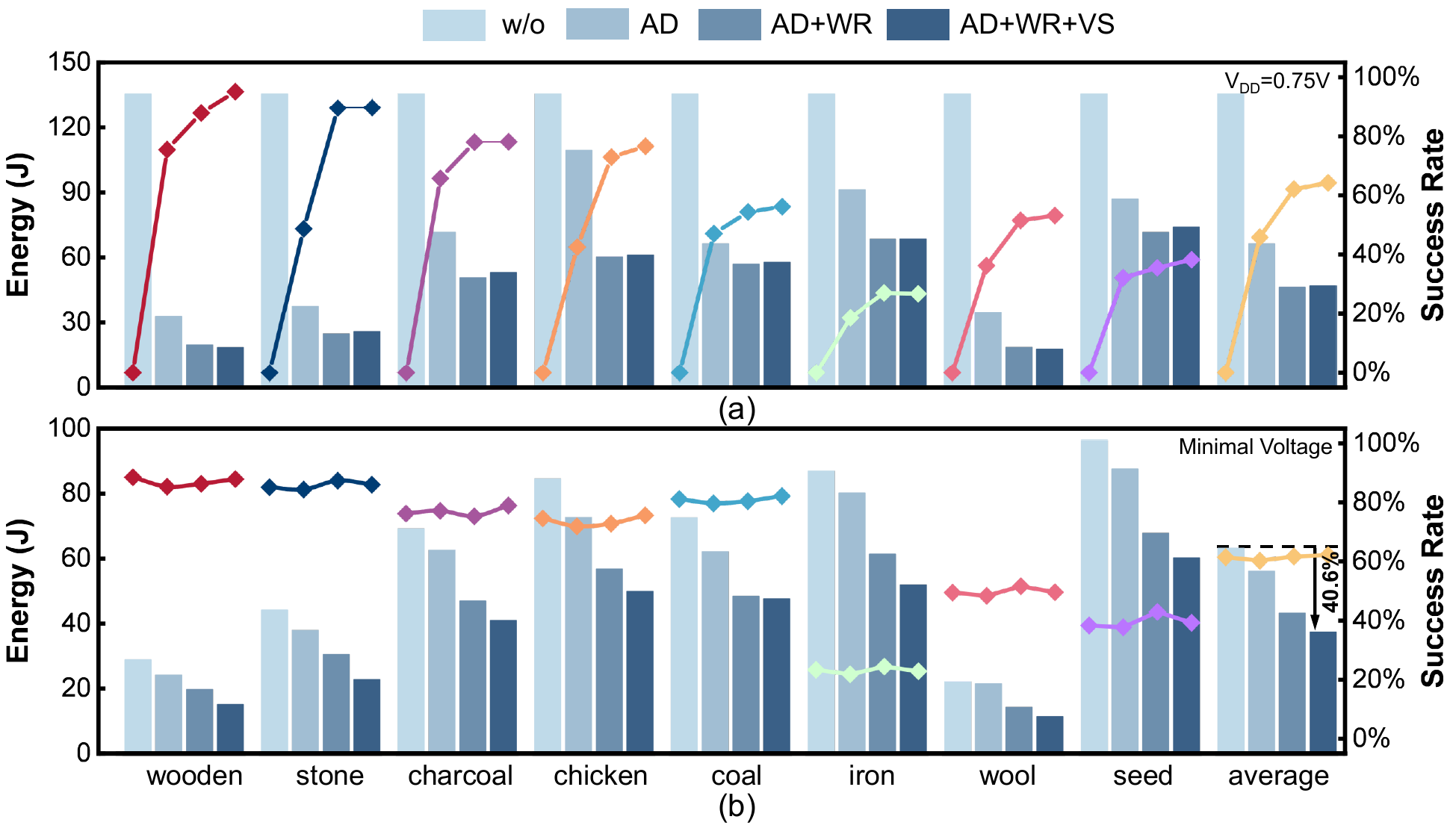}
    \vspace{-15pt}
    \caption{\textbf{Overall evaluation across different tasks.} (a) At 0.75\,V, \method~maintains success rates while reducing execution steps and energy consumption. (b) By supporting aggressive voltage scaling, \method~achieves an average 40.6\% energy savings without performance degradation.}
    \label{fig: overall evaluation across tasks} 
    \vspace{-5pt}
\end{figure}

\begin{figure}[!tb]
    \centering
    \includegraphics[width=1\linewidth]{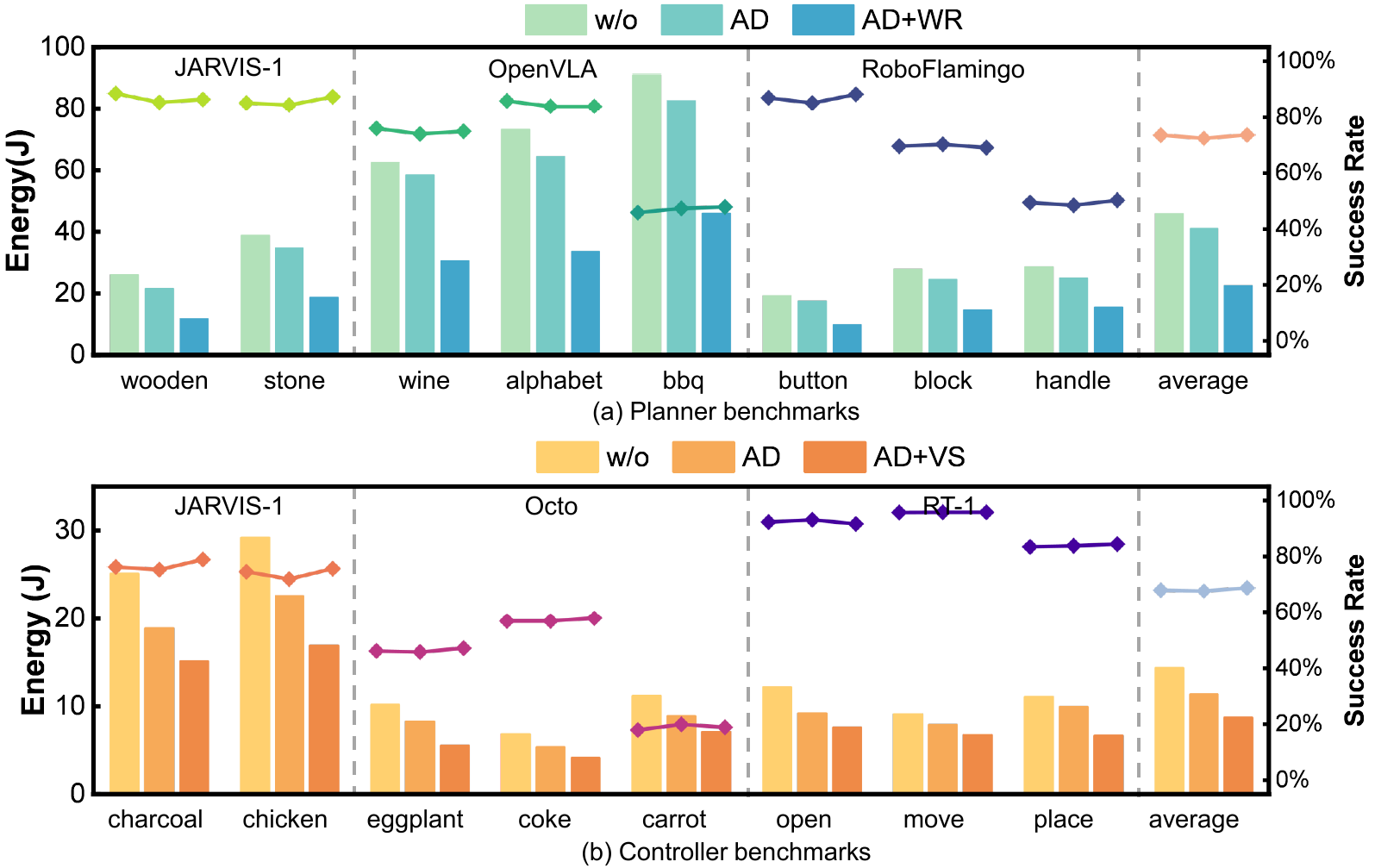}
    \caption{\textbf{Cross-platform evaluation.} Across various platforms and tasks, our techniques consistently achieve energy savings. AD+WR reduces planner energy consumption by 50.7\% on average, while AD+VS achieves 39.3\% average power savings for controllers. } 
    \label{fig: cross_platform} 
    \vspace{-10pt}
\end{figure}

\textit{\textbf{Reliability Improvement.}}
Fig.~\ref{fig: overall evaluation across tasks}(a) compares success rates and total computational energy consumption across four configurations and eight tasks: no protection, AD only, AD+WR, and AD+WR+VS. 
Operating at 0.75V without protection leads to near-zero success rates.
AD alone recovers an average of 71\% of the error-free success rate, while combining AD with WR boosts this reliability to 97\%, closely approaching the baseline. 
VS, though not designed to enhance fault tolerance, adds minimal overhead.
By improving task quality, total energy consumption is significantly reduced through fewer average steps.

\textit{\textbf{Efficiency Improvement.}}
Fig.~\ref {fig: overall evaluation across tasks}(b) compares the \xietong{computational} energy consumption under the lowest voltages across tasks that preserves success rates.
Compared to nominal voltage scenarios, AD alone results in an average 11.1\% reduction in energy consumption. WR further allows much lower planner voltage, achieving 18.8\% computational energy savings. VS optimizes controller voltage dynamically, contributing an additional 10.7\% savings. These techniques collectively achieve 40.6\% energy savings without compromising task success rates.


\xietong{
\textit{\textbf{Cross-platform Generality Evaluation}}.
To further validate the generality of our proposed method, we apply our techniques to additional representative embodied AI platforms. Due to limited open-source support for robust end-to-end robotic platforms that integrate both planners and controllers, we evaluate energy savings separately. For planners, we apply AD+WR to OpenVLA  (\texttt{wine}, \texttt{alphabet}, \texttt{bbq} on LIBERO) and RoboFlamingo  (\texttt{button}, \texttt{block}, \texttt{handle} on CALVIN). For controllers, we employ AD+VS onto Octo (\texttt{eggplant}, \texttt{coke}, \texttt{carrot} on OXE) and RT-1 (\texttt{open}, \texttt{move}, \texttt{place} on OXE). Detailed task descriptions are provided in the Appendix \ref{appendix: task}. Fig.~\ref{fig: cross_platform} shows the resulting computational energy savings while maintaining task quality. Across all platforms and tasks, our techniques consistently reduce energy consumption. AD+WR reduces planner energy consumption by 50.7\% on average, while AD+VS achieves 39.3\% average savings for controllers. 
}

\begin{figure}[!tb]
    \centering
    \includegraphics[width=1.0\linewidth]{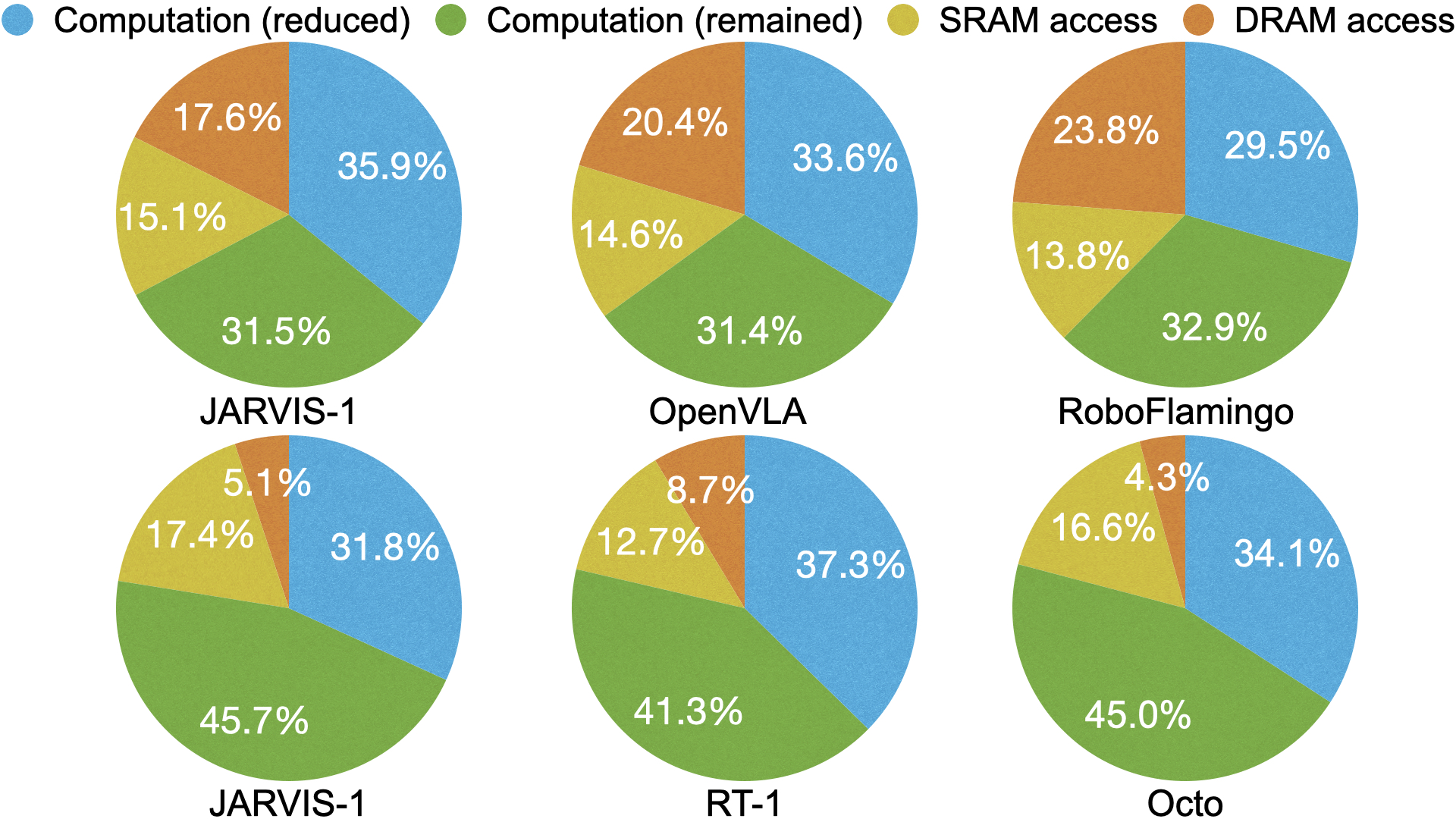}
    \vspace{-5pt}
    \caption{\textbf{Chip-level energy breakdown.} For planners, computation accounts for about 65\% of total energy and AD+WR reduces the computation energy by about 55\%, resulting in about 30\% total energy savings. For controllers, computation dominates over 75\% of total energy, and AD+VS achieves about 45\% reduction in computation energy, translating to about 35\% total energy savings.
        }
    \label{fig: energy breakdown} 
    \vspace{-8pt}
\end{figure}

\subsection{{Energy Consumption Breakdown and Analysis}}
\label{sec: energy}
\xietong{Sec. \ref {sec: multi-task} demonstrates that our \method~framework can consistently improve the reliability and efficiency of a wide range of embodied AI platforms.
In this subsection, we analyze how these computational energy savings translate to overall chip-level energy reduction and extended battery life for the entire robotic system.

The chip-level energy consumption can be broadly divided into two components: computation and memory access, with the latter including both on-chip SRAM and off-chip DRAM accesses. Table~\ref{tab:model_stats} summarizes the total number of parameters and computations for all models evaluated in this work. Energy estimates are derived by combining SCALE-Sim simulation results with post-layout synthesis reports and HBM2 specifications \cite{jun2017hbm}. 
While planner weights require off-chip loading for each inference, the controller weights can be stored entirely within on-chip SRAM, avoiding repeated off-chip accesses during execution. 
}

\begin{table}[ht]
\centering
\caption{Model parameters and computational requirements}
\vspace{-5pt}
\label{tab:model_stats}
\renewcommand*{\arraystretch}{1.05}
\begin{threeparttable} 
\resizebox{\linewidth}{!}{
\begin{tabular}{c|cccc}
\hline \hline
\textbf{Model} & \textbf{\#Params (M)} & \textbf{Input} & \textbf{Output}   & \textbf{GOps\tnote{3}}  \\
\hline \hline
JARVIS-1 planner \cite{wang2024jarvis} & 7,869 &740\tnote{1} & 25\tnote{1} & 5,344 \\ 
OpenVLA \cite{kim2024openvla} & 6,929 & 617\tnote{1} & 7\tnote{1} & 4,595  \\
RoboFlamingo \cite{li2023vision} & 2,552 & 505\tnote{1} & 6\tnote{1} & 2,411 \\ \hline
JARVIS-1 controller \cite{lifshitz2023steve} & 61 &128\tnote{2} &-& 102 \\
RT-1 \cite{brohan2022rt} & 35 &224\tnote{2} &-& 78 \\
Octo \cite{ghosh2024octo} & 27 &224\tnote{2} &-& 76\\ \hline
Entropy predictor & 0.055 &- &- &0.043\\ 
\hline \hline
\end{tabular}
}
\begin{tablenotes} 
\footnotesize
\item \tnote{1}The input and output sizes of the planner correspond to the \\ representative number of prefill and decode tokens, which may \\vary across executions. \tnote{2}For controllers, the input size refers to \\ the resolution of square RGB images. \tnote{3}Giga operations @ INT8.
\end{tablenotes}
\end{threeparttable}
\vspace{-5pt}
\end{table}

\xietong{
Fig.~\ref{fig: energy breakdown} shows the energy breakdown for each planner and controller. Computation accounts for 67.4\%, 65.0\%, and 62.4\% of total energy in the planners, and 77.5\%, 78.6\%, and 79.1\% in the controllers (due to amortized DRAM overhead). Consequently, the gains profiled in Sec. \ref{sec: multi-task} translate to chip-level energy savings of 35.9\%, 33.6\%, and 29.5\% for the planners, and 31.8\%, 37.3\%, and 34.1\% for the controllers, respectively.

Although some savings may be partially offset by energy consumption from mechanical components (e.g., actuators and motors), we estimate an overall battery life extension of 15–30\% in realistic robotic deployments according to several popular configurations \cite{huang2025dadu, jabbour2024generative, anybotics2023anymal, kim2024openvla, gaz2019dynamic}, 
where computation accounts for energy consumption comparable to or exceeding that of mechanical components.
These gains come with negligible impact on task success, computational latency, or chip area, making the proposed approach a practical and meaningful optimization for real-world embodied AI deployment. Besides, this computational power reduction would also ease heat dissipation issues.

}

\subsection{{Configuration Analysis}}
\label{sec: configuration}

\begin{figure}[!tb]
    \centering
    \includegraphics[width=1\linewidth]{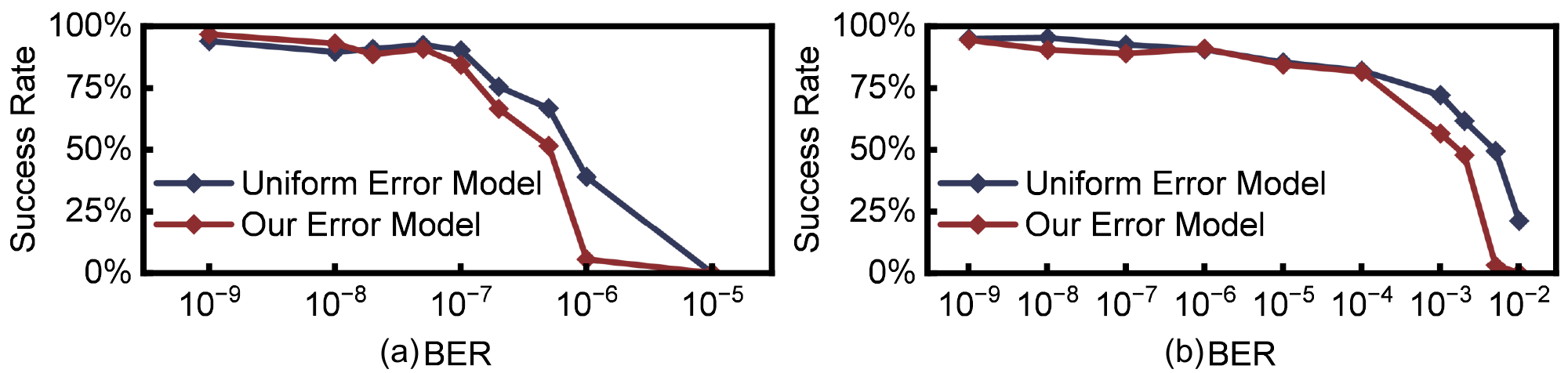}
    \vspace{-10pt}
    \caption{\textbf{Comparison of uniform and hardware-specific error models} applied to (a) the planner and (b) the controller on the \texttt{wooden} task in JARVIS-1. 
        }
    \label{fig: Error model validity} 
\end{figure}

\xietong{
This subsection investigates the impact of several manually defined configurations. First, we validate the conclusions drawn in Sec. \ref{sec: resilience characterization} under uniform error model. Next, we analyze how the number of experiment repetitions affects the measured task performance. Finally, we discuss the implications of applying more aggressive quantization, such as INT4.

\textbf{\textit{Validity of Resilience Characterization}}.
In Sec.\ref{sec: resilience characterization}, we adopt a uniform error model to derive generalizable algorithmic insights, while in Sec. \ref{sec: experiment}, error model profiled in Fig. \ref{fig:error_model} is used for more accurate power estimation under voltage scaling. Fig.~\ref{fig: Error model validity} presents success rates for the \texttt{wooden} task, showing the impact of both error models. Although there are slight numerical differences, the overall performance trends remain consistent, validating that resilience characterization is independent of the specific error model.

\textbf{\textit{Statistical significance of repetitions}}.
In this work, we repeat our experiments for \textit{at least} 100 times to guarantee a 95\% confidence interval of 3\% to 5\%. This is sufficient to reveal an overall trend that supports our conclusions. Table~\ref{tab: repetitions} presents the measured success rate of the \texttt{wooden} task as the number of repeated experiments increases. As shown, 100 repetitions have been sufficient to get converged results.
}

\begin{table}[!tb]
    \centering
    \caption{Measured success rates versus repetitions.} 
    \label{tab: repetitions}  
    \vspace{-7pt}
    \renewcommand{\arraystretch}{1.05}
    \resizebox{\linewidth}{!}{
    \begin{tabular}{c|c c c c c c} 
        \hline\hline
        \# Repetitions & 20   & 40   & 60    & 80   & 90   & 100  \\
        \hline
        Success Rate & 85.0\% & 90.0\% & 91.7\%  & 92.5\% & 91.1\% & 90.0\% \\
        \hline\hline
        \# Repetitions & 110  & 120  & 140   & 160  & 180  & 200  \\
        \hline
        Success Rate  & 89.1\% & 89.2\% & 90.7\%  & 90.6\% & 89.4\% & 89.5\% \\
        \hline\hline  
        \multicolumn{7}{l}{\footnotesize Measured on \texttt{wooden} task, BER = $1\times 10^{-7}$ injected to the JARVIS-1 controller.} \\
    \end{tabular}%
    }
\end{table}


\textbf{\textit{Quantization-related behaviors}}.
Although this work chooses INT8 as a default configuration, more aggressive quantization (e.g., INT4) offers an orthogonal approach to reduce power consumption and extend battery life, since it decreases not only computational complexity but also memory overhead. At the circuit level, when quantized to INT4, timing errors in higher bits would still be frequent since their critical paths are longer. At the algorithm level, we present the effects of AD+WR for INT8 and INT4 in Table \ref{tab: quantization behavior}. 
The current results are not statistically different.
We hypothesize that this is because, compared to INT8, INT4 reduces the error-free success rate due to larger quantization error. However, after applying AD+WR, the undetected error range below the detection threshold is compressed, thereby enhancing robustness against injected errors.


\begin{table}[!tb]
    \centering
    \caption{Success rate on \texttt{stone} when applying AD+WR}
    \label{tab: quantization behavior}   \vspace{-7pt}
    \renewcommand*{\arraystretch}{1.05}
    \resizebox{\linewidth}{!}{
    \begin{tabular}{ c | c c c c c } 
        \hline\hline
        BER & $1\times10^{-6}$ & $1\times10^{-5}$ & $1\times10^{-4}$ & $1\times10^{-3}$ & $1\times10^{-2}$ \\
        \hline
        INT8 & 90.4\% & 91.5\% & 91.4\% & 90.5\% & 85.6\% \\
        INT4 & 91.0\% & 88.3\% & 88.2\% & 89.5\% & 87.7\% \\
        \hline\hline  
    \end{tabular}
    }
\end{table}

\subsection{Comparison with Existing Techniques}
\label{sec: existing techniques}
\begin{figure}[!tb]
    \centering
    \includegraphics[width=1.0\linewidth]{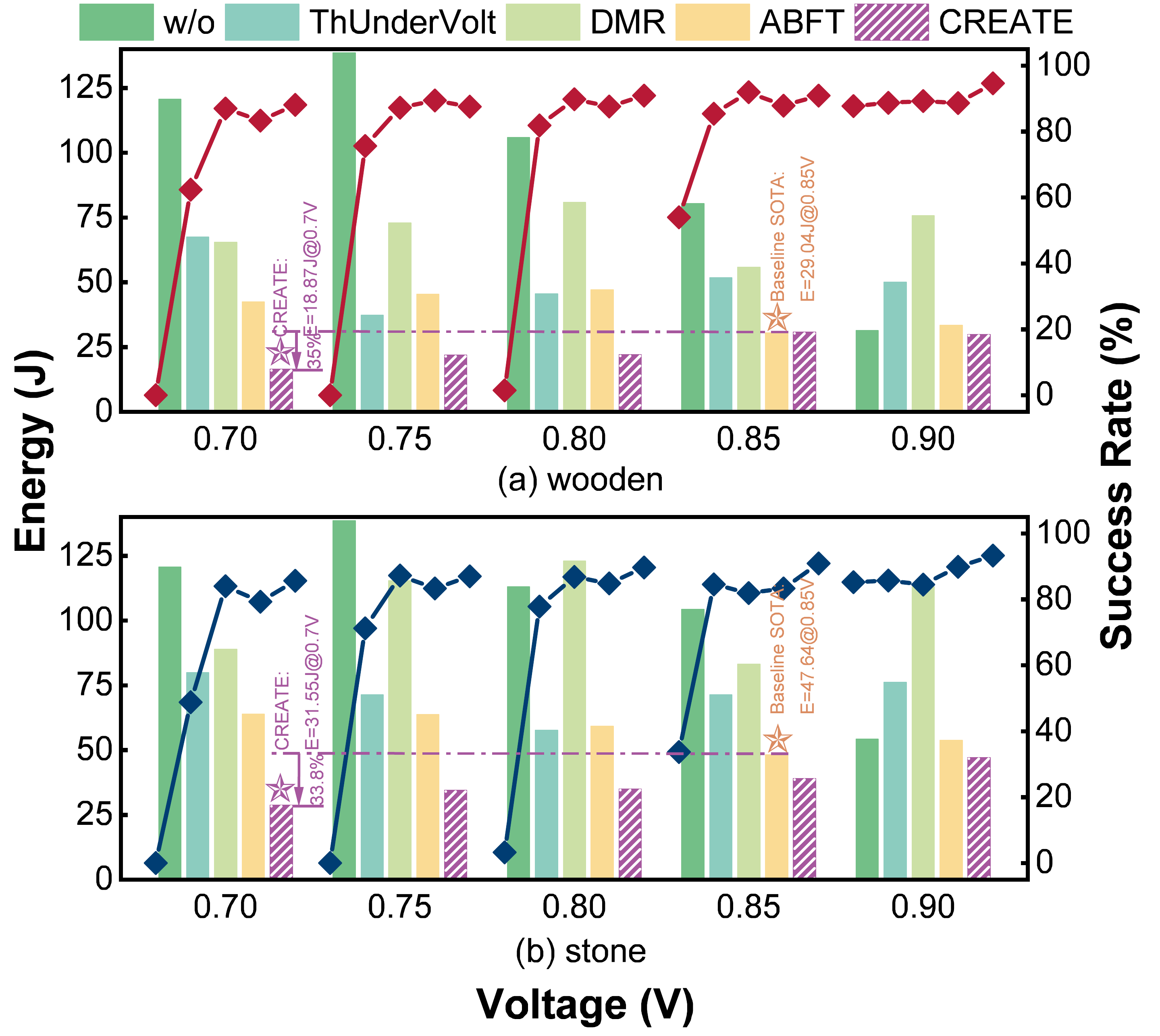}
    \vspace{-5pt}
    \caption{\textbf{Existing technique comparisons}.  \method~ consistently outperforms all baselines across operating voltages, reducing energy consumption by (a) 35.0\% for \texttt{wooden} and (b) 33.8\% for \texttt{stone} over SOTA baseline.}
    \label{fig: comparison with baseline} 
    \vspace{-5pt}
\end{figure}

In this subsection, we compare \method~with existing compatible circuit- and algorithm-level baselines. On the circuit level, we select a representative timing-borrowing approach, ThUnderVolt \cite{zhang2018thundervolt}, and widely used DMR \cite{talpes2020compute}; on the algorithm level, we adopt SOTA ABFT variants \cite{xue2023approxabft}. Fig.~\ref{fig: comparison with baseline} shows success rates and total energy consumption across various operating voltages. 
\xietong{
    DMR achieves high reliability but incurs at least 2$\times$ overhead and frequent error recoveries, bringing about prohibitive energy cost. ThUnderVolt adds bypass circuits to all PEs in order to skip faulty results with slightly lower circuit overhead, but because of excessive neuron pruning, it significantly degrades performance and increases execution steps that ultimately increase power consumption at low supply voltages. ABFT uses lightweight checksums to correct errors, but its application below 0.85V is confined because of frequent error recovery overhead as BER increases. In contrast, \method~introduces negligible circuit overhead, avoids recovery mechanisms, and maintains task quality, achieving 35.0\% and 33.8\% energy savings for \texttt{wooden} and \texttt{stone}, respectively.}


\vspace{-5pt}
\section{Conclusion}

We introduce \method, a 
\xietong{general design principle that exploits the heterogeneous resilience across circuit-algorithm-application layers} to synergistically enable efficient yet reliable embodied AI systems. For the first time, we perform extensive error injection experiments to characterize the resilience of embodied AI systems. Our results highlight that the planner exhibits a much worse resilience than the controller due to systematic outliers in the activation distribution, while the controller's resilience exhibits a dependency on subtasks and execution steps.  Building on these insights, \method~co-optimizes efficiency and reliability across circuit, model, and application layers, with a voltage scaling architecture design.
\xietong{By enabling aggressive voltage reduction, CREATE consistently achieves energy savings across diverse embodied AI platforms and tasks. It reduces computational energy consumption by an average of 40.6\% compared to nominal-voltage operation and 35.0\% over baseline SOTA methods, which translates to 29.5\% to 37.3\% chip-level energy savings. By facilitating energy-efficient embodied AI deployment, our framework extends overall system battery life by about 15\% to 30\%, broadening the potential application scope for embodied AI systems.}


\begin{acks}
This work was supported in part by the National Natural Science Foundation of China (NSFC) under Grant 62125401, Grant 62495102, Grant 92464104, and Grant 62341407; in part by the National Key Research and Development Program under Grant 2024YFB4505004; in part by the Beijing Municipal Science and Technology Program under Grant Z241100004224015; in part by the Beijing Outstanding Young Scientist Program under Grant JWZQ20240101004; and in part by the 111 Project under Grant B18001.
\end{acks}


\newpage
\bibliographystyle{ieeetr}
\bibliography{reference}
\appendix
\section{Model Details}
\label{appendix: model}

This work involves multiple models, whose detailed architecture, such as the number of layers, hidden dimensions, parameters, and FLOPs, are critical hyperparameters. Table~\ref {tab: planners} to \ref{tab: entropy_predictor_architecture} presents the detailed architectures of the planner, controller, and our proposed entropy predictor. 
Table~\ref {tab:model_stats} further summarizes the total number of parameters and computations for each model. 
For brevity, only primary modules are listed, with some secondary components omitted. The kernel size of Conv2d is 3 by default.
\vspace{-10pt}

\begin{table}[htb]
    \centering
    \caption{Model architecture of planners}  
    \vspace{-10pt}
    \label{tab: planners}   
    \renewcommand*{\arraystretch}{1.05}
    \resizebox{0.9\linewidth}{!}{  
    \begin{tabular}{ c |c c c } 
        \hline\hline
        \textbf{Model} & \textbf{\#Layers} & \textbf{Hidden Dim} & \textbf{MLP Dim} \\
        \hline \hline
        JARVIS-1     & 32    & 4,096     & 14,336   \\
        OpenVLA      & 32    & 4,096     & 11,008   \\
        RoboFlamingo & 24    & 2,048     & 8,192    \\
        \hline\hline  
    \end{tabular}
    }
    \vspace{-20pt}
\end{table}

\begin{table}[ht]
\centering
\caption{Model architecture of controllers}
\vspace{-10pt}
\label{tab: controllers}
\renewcommand*{\arraystretch}{1.05}
    \resizebox{1.0\linewidth}{!}{
\begin{tabular}{c|ccccc}
\hline \hline
\textbf{Model} & \textbf{Parts}     & \textbf{\# Layers} & \textbf{Hidden Dim} & \textbf{MLP Dim} \\ \hline \hline
\multirow{2}{*}{JARVIS-1}        & Img$^*$     & 10                 & 3-256                 & 128-256                 \\ 
                & Decoder           & 4                  & 1,024                  & 4,096                   \\ \hline
    \multirow{3}{*}{RT-1}              & MaxViT            & 11                 & 64-768                & 384-3,072               \\  
          & FiLM              & 23                 & 1,536                  & 256-3,072               \\ 
                & ViT               & 6                  & 768                   & 3,072                   \\ \hline
                & T5                & 12                 & 384                   & 1,152                   \\ 
Octo            & Obs$^*$        & 8                  & 6-384                 & 32-512                 \\ 
                & Head              & 3                  & 256                   & 1,024                   \\ \hline \hline
\multicolumn{5}{@{}l@{}}{\footnotesize $^*$Conv2d layers, where the hidden and MLP dim refer to input and output channels.} \\
\end{tabular}
\vspace{-20pt}
}

\end{table}

\vspace{-15pt}
\begin{table}[htp]
\centering
\renewcommand*{\arraystretch}{1.05}
    \caption{Model architecture of the entropy predictor}
    \vspace{-5pt}
    \label{tab: entropy_predictor_architecture}
    \resizebox{\linewidth}{!}{
\begin{tabular}{c|c|c}
\hline \hline
\textbf{Block} & \textbf{Layer Type} & \textbf{Configuration} \\
\hline \hline
\multirow{9}{*}{\makecell{Image \\ Processing \\ (CNN)}} 
    & Conv2d    & in=3, out=16, stride=3, padding=1  \\
    & ReLU      &                                     \\
    & MaxPool2d & kernel\_size=2                      \\
    & Conv2d    & in=16, out=32, stride=3, padding=1 \\
    & ReLU      &                                     \\
    & MaxPool2d & kernel\_size=2                      \\
    & Conv2d    & in=32, out=64, stride=3, padding=1 \\
    & ReLU      &                                     \\
    & AvgPool2d & output\_size=1                      \\
\hline
\multirow{1}{*}{\makecell{Prompt MLP}} 
    & Linear    & in=512, out=64                     \\
\hline
\multirow{4}{*}{Fusion MLP}
    & Concat    &                                     \\
    & Linear    & in=128, out=128                     \\
    & ReLU      &                                     \\
    & Linear    & in=128, out=1                       \\
\hline \hline
\end{tabular}
}
\end{table}

\section{Task Description}
\label{appendix: task}

This paper evaluates 21 tasks across four benchmarks, which are referred to as abbreviations throughout this paper for brevity. Detailed task descriptions are provided in Table~\ref{tab:benchmark_tasks}.

\begin{table}[H]
    \centering
    \caption{Task descriptions}
    \vspace{-10pt}
    \label{tab:benchmark_tasks}
    \renewcommand*{\arraystretch}{1.05}
    \resizebox{1.0\linewidth}{!}{
    \begin{tabular}{ccc}  
        \hline\hline
        \textbf{Benchmark} & \textbf{Abbr.} & \textbf{Task Description} \\
        \hline\hline
        \multirow{9}{*}{Minecraft}  
            & \texttt{wooden}    & Obtain a wooden pickaxe in a jungle \\
            & \texttt{stone}     & Obtain a stone pickaxe in the plains \\
            & \texttt{charcoal}  & Obtain charcoal in the plains \\
            & \texttt{chicken}   & Obtain a cooked chicken in the plains \\
            & \texttt{coal}      & Obtain coal in a savanna \\
            & \texttt{iron}      & Obtain an iron sword in the plains \\
            & \texttt{wool}      & Obtain 5 white wool in the plains \\
            & \texttt{seed}      & Obtain 10 wheat seeds in a savanna \\
            & \texttt{log}      & Obtain 10 logs in a forest \\
        \hline
        \multirow{3}{*}{LIBERO}
            & \texttt{wine}      & Put wine bottle on top of cabinet \\
            & \texttt{alphabet}  & Pick up alphabet soup and place it in basket \\
            & \texttt{bbq}       & Pick up bbq sauce and place it in basket \\
        \hline
        \multirow{3}{*}{CALVIN}
            & \texttt{button}    & Press the button to turn off the LED light \\
            & \texttt{block}     & Slide the block that it falls into the drawer \\  
            & \texttt{handle}    & Pull the handle to open the drawer \\
        \hline
        \multirow{6}{*}{OXE}
            & \texttt{eggplant}  & Put eggplant in basket \\
            & \texttt{coke}      & Grasp single opened coke can \\  
            & \texttt{carrot}    & Put carrot on plate \\
            & \texttt{open}      & Open middle drawer  \\
            & \texttt{move}      & Move near google baked tex  \\  
            & \texttt{place}     & Place into closed top drawer  \\
        \hline\hline
    \end{tabular}
    }
\end{table}

\vspace{-10pt}
\section{Entropy-to-Voltage Mapping Policy}

\begin{figure}[htb]
    \centering
    \includegraphics[width=1\linewidth]{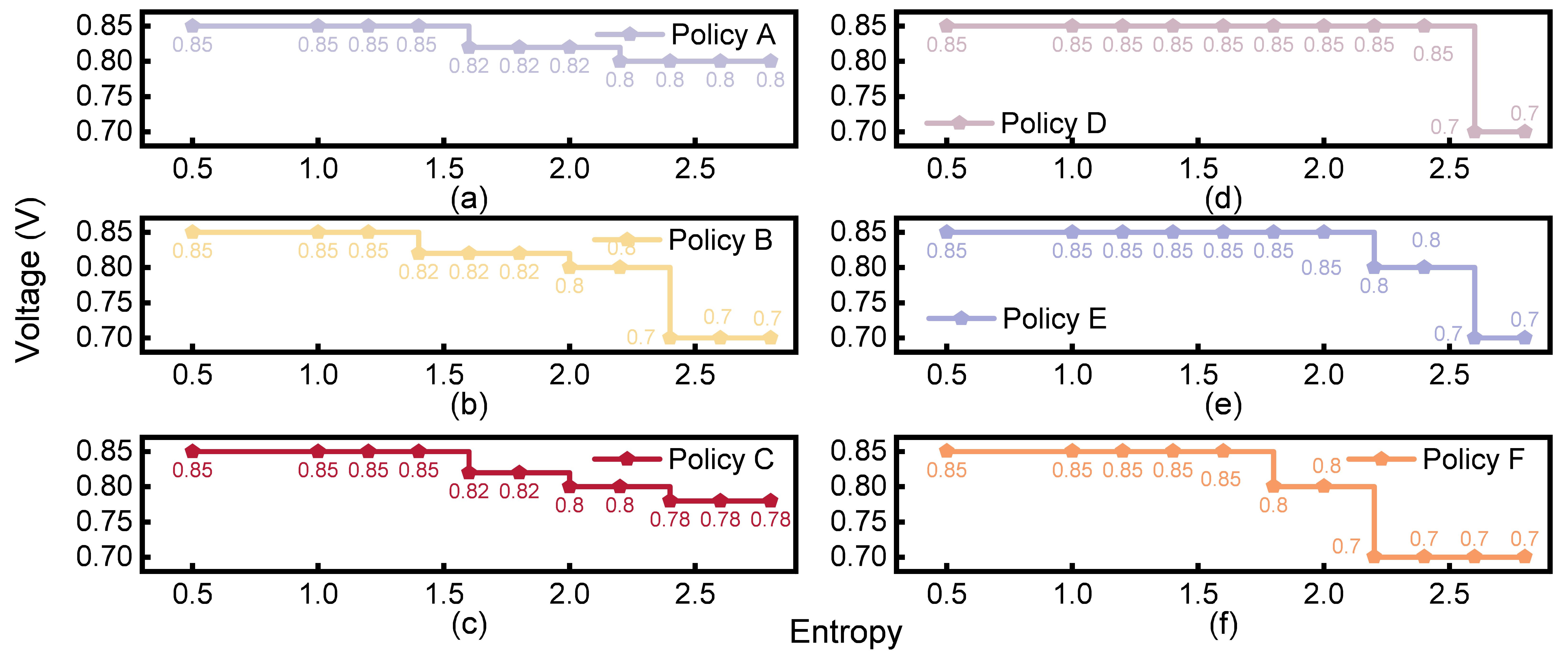}
    \caption{Selected entropy to voltage mapping policies} 
    \label{fig:entropy-to-voltage} 
\end{figure}

Fig. \ref{fig:entropy-to-voltage} illustrates the detailed entropy-to-voltage mapping policies A to F for JARVIS-1, which are searched from over 100 candidates.

\end{document}